\documentclass[12pt]{article}
\usepackage{latexsym}
\usepackage{fontenc}
\usepackage{amssymb}
\usepackage{amsmath}
\usepackage[all]{xy}
\usepackage[dvipdf]{epsfig}
\usepackage{color}
\usepackage{graphicx}
\usepackage{amsfonts}
\numberwithin{equation}{section}

\begin{document}
\title{\begin{flushright}
\normalsize{MZ-TH/09-17}
\bigskip
\end{flushright}
Bimetric Truncations for \\
Quantum Einstein Gravity and \\
Asymptotic Safety}
\date{}
\author{Elisa Manrique 
and Martin Reuter\\
Institute of Physics, University of Mainz\\
 Staudingerweg 7, D-55099 Mainz, Germany}
\maketitle

\begin{abstract} 
In the average action approach to the quantization of gravity the fundamental requirement of ``background independence"
is met by actually introducing a background metric but leaving it completely arbitrary. The associated Wilsonian
renormalization group defines a coarse graining flow on a theory space of functionals which, besides the dynamical
metric, depend explicitly on the background metric. All  solutions to the truncated flow equations known to date have 
a trivial background field dependence only, namely via the classical gauge fixing term. In this paper we analyze a number of conceptual
issues related to the bimetric character of the gravitational average action and explore a first nontrivial
bimetric truncation in the simplified setting of conformally reduced gravity. Possible implications for the Asymptotic Safety 
program and the cosmological constant problem are discussed in detail.

\end{abstract}

\bigskip

\section{Introduction and motivation}\label{section1}

Unifying the principles of quantum mechanics and general relativity is perhaps still the most challenging
open problem in fundamental physics \cite{kiefer}. While the various approaches that are currently being
developed, such as string theory, loop quantum gravity \cite{A,R,T}, or asymptotic safety \cite{wein}-\cite{livrev},
for instance, are based upon rather different physical ideas and are formulated in correspondingly different
mathematical frameworks, they all must cope with the problem of ``background independence" in one way or another.
Whatever the ultimate theory of quantum gravity will look like, a central requirement we impose on it is that it should be 
``background independent" in the same sense as general relativity is background independent. Loosely speaking, this means that
the spacetime structure actually realized in Nature should not be part of the theory's definition but rather emerge
as a solution to certain dynamical equations. In classical general relativity the spacetime structure is encoded
in a Lorentzian metric on a smooth manifold, and this metric, via Einstein's equation, is a dynamical consequence of  the matter
present in the Universe.

\subsection{The requirement of ``background independence"}\label{subsection1.1}

In the following we shall explore the possibility of constructing a quantum field theory of gravity in which the metric
carries the dynamical degrees of freedom. Even though this property is taken over from classical general relativity the
fundamental dynamics of those metric degrees of freedom, henceforth denoted $\gamma_{\mu\nu}$, is allowed to be different
from that in classical general relativity.

Furthermore, the theory of quantum gravity we are searching for will be required to respect the principle of ``background independence":
In the formulation of the theory no special metric should play any distinguished role. The actual metric of spacetime should 
arise as the expectation value of the quantum field (operator) $\gamma_{\mu\nu}$ with respect to some state: 
$g_{\mu\nu}=\langle\gamma_{\mu\nu}\rangle$. This is in sharp contradistinction to the traditional quantum field theory on Minkowski
space whose conceptual foundations heavily rely on the availability of a non-dynamical (rigid) Minkowski spacetime as a background structure.

Trying to set up a similar quantum field theory of the metric itself, let us assume that we are given some candidate for a microscopic
interaction which is described by a diffeomorphism invariant classical (i.e. bare) action functional $S[\gamma_{\mu\nu}]$.
Whatever this action is (Einstein-Hilbert, $R^2,\cdots$, etc.), well before one encounters the notorious problems related to the
ultraviolet (UV) divergences, profound conceptual problems arise. In absence of a rigid background when the metric is dynamical, 
there is no preferred time direction, for instance, hence no notion of equal time commutators, and clearly the usual rules of quantization 
cannot be applied straightforwardly.
Many more problems arise when one tries to apply the familiar methods of quantum field theory to the metric itself without introducing
a rigid background structure. Some of them are conceptually deep, others are of a more technical nature. Here we only mention one type
of difficulties which later on will become central in our discussion. 

In conventional field theory on a rigid background typical 
regulator schemes (by higher derivatives, for example) which are used to make the theory well behaved, both in the infrared (IR)
and the UV, employ the background metric. As a result, it is not obvious if and how such schemes can be applied to quantum gravity.
This problem is particularly acute for approaches based upon some sort of functional renormalization group equation (FRGE) which is
supposed to implement a Wilson-like renormalization group (RG) flow by a continuous coarse graining  \cite{avact}-\cite{ymrev}.
In conventional Euclidean field theory and statistical mechanics every such coarse graining is described by an associated length scale
which measures the size of the spacetime blocks within  which the microscopic degrees of freedom got averaged over. When the metric is
dynamical and no rigid background is available, this concept becomes highly problematic since it is not clear a priori in terms of which
metric one should express the physical, i.e. proper diameter of a spacetime block.

The principle of ``background independence"\footnote{Here and in the following we use quotation marks when ``background independence" is
supposed to stand for this principle (rather than for the independence of some quantity of the background field).} which we would like to 
implement in the quantum field theory of the metric we are aiming at can be summarized as the requirement that 
{\it none of the theory's basic rules and assumptions, calculational methods, and none of its predictions, therefore, may depend on any
special metric fixed a priori. All metrics of physical relevance must result from the intrinsic quantum gravitational dynamics.}

A possible objection\footnote{This argument is due to D. Giulini \cite{giulini}. We thank him and A. Ashtekar for a discussion of this point.}
against this working definition could be as follows: A theory can be made ``background independent" in the above sense, but 
nevertheless has a distinguished rigid background if the latter arises as the unique solution to some field equation which is
made part of the ``basic rules". For instance, rather than introducing a Minkowski background directly one simply
imposes the field equation $R^{\mu}_{\phantom{1}\nu\rho\sigma}=0$. However, this objection can apply only in a setting where the dynamics,
the field equations, can be chosen freely. In asymptotically safe gravity, the case we are actually interested in, this
is impossible as the dynamics is dictated by the fixed point action.

There are two quite different strategies for complying with the requirement of ``background independence": 
{\bf (i)} One can try to define the theory, and work out its implications, without ever employing a background metric or  a similar 
nondynamical structure. While this is the path taken in loop quantum gravity \cite{A,R,T} and the discrete approaches
\cite{hamber},\cite{ajl1}-\cite{ajl3}, for instance, it seems very hard , if not impossible to realize it in a continuum field
theory.\footnote{The typical difficulties are reminiscent of those encountered in the quantization of topological Yang-Mills theories.
Even when the classical  action can be written down without the need of a metric, the gauge fixing and quantization 
of the theory usually requires one.
Hence the only way of proving  the topological character of some result is to show its independence of the metric chosen.} 
{\bf (ii)} One employs an arbitrarily chosen background metric $\bar{g}_{\mu\nu}$ at the intermediate steps of the quantization,
but verifies at the end that no physical prediction depends on which metric was chosen. This is the route taken in the 
gravitational average action approach \cite{mr} which we are going to use in this paper.

\subsection{The bimetric solution to the ``background independence" problem}\label{subsection1.2}

In the average action approach one decomposes the quantum metric as $\gamma_{\mu\nu}=\bar{g}_{\mu\nu}+h_{\mu\nu}$ and quantizes the 
(non-linear) fluctuation $h_{\mu\nu}$ in essentially the same way one would quantize a matter field in a classical spacetime
with metric $\bar{g}_{\mu\nu}$. In this way all of the conceptual problems alluded to above, in particular the difficulties related 
to the construction of regulators, disappear. Technically the quantization of gravity proceeds then almost as in standard field
theory on a rigid classical spacetime, with one essential difference, though: In the latter, one concretely fixes the background 
$\bar{g}_{\mu\nu}$ typically as $\bar{g}_{\mu\nu}=\eta_{\mu\nu}$ or as $\bar{g}_{\mu\nu}=\delta_{\mu\nu}$ in the Euclidean case.
In ``background independent" quantum gravity instead, the metric $\bar{g}_{\mu\nu}$ is never concretely chosen. All objects that one
has to compute in this setting, generating functionals, say, are functionals of the variable $\bar{g}_{\mu\nu}$. An example is the 
effective action $\Gamma[\bar{h}_{\mu\nu};\bar{g}_{\mu\nu}]$ which depends on the background and the fluctuation expectation value
$\bar{h}_{\mu\nu}\equiv\langle h_{\mu\nu}\rangle$. In a sense, {\it the ``background independent" quantization of gravity amounts to its
quantization on all possible backgrounds at a time}.

There are two metrics now which are almost equally important: the background $\bar{g}_{\mu\nu}$ and the expectation value metric
\begin{equation}\label{1.1}
g_{\mu\nu}\equiv\langle \gamma_{\mu\nu}\rangle=\bar{g}_{\mu\nu}+\bar{h}_{\mu\nu},\qquad \bar{h}_{\mu\nu}\equiv\langle h_{\mu\nu}\rangle
\end{equation}
Alternatively we may regard the effective action as a functional of the two metrics, defining
\begin{equation}\label{1.2}
\Gamma[g_{\mu\nu},\bar{g}_{\mu\nu}]\equiv \Gamma[\bar{h}_{\mu\nu}=g_{\mu\nu}-\bar{g}_{\mu\nu}\; ;\; \bar{g}_{\mu\nu}]
\end{equation}
In this language both arguments of $\Gamma$ can be varied freely over the same space of tensor valued functions.
Because of the symmetric status enjoyed by the two metrics we can characterize this setting as a ``bimetric" approach.

\subsection{Background-covariant coarse graining}\label{subsection1.3}

The effective average action (EAA) is a scale dependent version of the ordinary effective action, with built-in
IR cutoff at a variable momentum scale $k$ \cite{avact,avactrev,ymrev}. In the case of gravity \cite{mr} its
formal definition starts out from the modified gauge fixed path integral over $h_{\mu\nu}$ and the Faddeev-Popov
ghosts $C^{\mu}$ and $\bar{C}_{\mu}$:
\begin{equation}\label{1.3}
\int\mathcal{D}h_{\mu\nu}\mathcal{D}C^{\mu}\mathcal{D}\bar{C}_{\mu}\, \exp{\Big(-\widetilde{S}-\Delta_kS\Big)}
\end{equation}
Here $\widetilde{S}\equiv S+S_{\textrm{gf}}+S_{\textrm{gh}}$ includes, besides the bare action $S$, the gauge fixing and the ghost
terms. The new ingredient is the cutoff action which suppresses the IR modes. It has the structure
\begin{equation}\label{1.4}
\Delta_kS[h;\bar{g}]\propto \int\textrm{d}^4x\,\sqrt{\bar{g}}\;h_{\mu\nu}\;\mathcal{R}_k[\bar{g}]^{\mu\nu\rho\sigma}\;h_{\rho\sigma}
\end{equation}
plus a similar term for the ghosts. The coarse graining kernel $\mathcal{R}_k[\bar{g}]$ is a functional of $\bar{g}_{\mu\nu}$.
In fact, the background field is fundamental for a covariant (in the sense explained below) coarse graining and for giving a 
proper (as opposed to a pure ``coordinate") meaning to the momentum scale $k$. When the integration variable $h_{\mu\nu}$ is
expanded in eigenmodes of the covariant Laplacian $-\bar{D}_{\mu}\bar{D}^{\mu}$ constructed from $\bar{g}_{\mu\nu}$, the
eigenvalue of the lowest mode integrated out unsuppressed has eigenvalue $k^2$.

For the discussion in the present paper it is important to note that $\Delta_k S$ is a functional of the two independent
fields $h_{\mu\nu}$ and $\bar{g}_{\mu\nu}$. Contrary to a classical action it is not just a functional of their sum
$\gamma_{\mu\nu}\equiv \bar{g}_{\mu\nu}+{h}_{\mu\nu}$. We say that $\Delta_k S$ has an 
{\it extra background field dependence}, that is, a dependence on  $\bar{g}_{\mu\nu}$ which does not combine with 
$h_{\mu\nu}$ to form the full metric $\gamma_{\mu\nu}$. (The same is also true for $S_{\textrm{gf}}+S_{\textrm{gh}}$.)
For a detailed discussion of this point, and the corresponding background-quantum field split symmetry, we refer to
\cite{creh1} and \cite{creh2}.

\subsection{Properties of the gravitational average action}\label{subsection1.4}

With a modified path integral \eqref{1.3} as the starting point, the remaining steps in the construction of the gravitational EAA 
proceed almost as in the case of the ordinary effective action: one introduces source  terms for $h_{\mu\nu}$ and for the
ghosts, defines a connected generating functional $W_k$, and introduces its Legendre transform 
$\widetilde{\Gamma}[\bar{h}_{\mu\nu},\xi^{\mu},\bar{\xi}_{\mu};\bar{g}_{\mu\nu}]$. It depends on the expectation values of  $h_{\mu\nu}$
and the Faddeev-Popov ghosts, denoted $\xi^{\mu}$ and $\bar{\xi}_{\mu}$, respectively. Finally, the EAA is defined as the
difference $\Gamma_k\equiv \widetilde{\Gamma}-\Delta_k S$ where the expectation value fields are inserted into $\Delta_k S$.

Let us list the main properties of the gravitational EAA, to the extent they are relevant to the present discussion.
For further details we refer to \cite{mr}.

\noindent
{\bf (A)} At $k=0$, $\Gamma_k$ coincides with the ordinary effective action $\Gamma\equiv\Gamma_0$.

\noindent
{\bf (B)} Originally, $\Gamma_k\equiv \Gamma_k[\bar{h},\xi,\bar{\xi};\bar{g}]$ depends, besides the background metric, on the expectation value
fields $\bar{h},\xi,\bar{\xi}$. This presentation of the EAA fits with the intuition that we are quantizing the three ``matter" fields
$h,C,\bar{C}$ in the classical $\bar{g}_{\mu\nu}$-background spacetime. Sometimes the bimetric point of view is more convenient.
Then one replaces $\bar{h}$ by $g\equiv \bar{g}+\bar{h}$ as the independent argument and defines
\begin{equation}\label{1.5}
\Gamma_k[g,\bar{g},\xi,\bar{\xi}]\equiv\Gamma_k[\bar{h}=g-\bar{g},\xi,\bar{\xi};\bar{g}]
\end{equation}
This notation allows us to interpret
 the second, i.e. the $\bar{g}_{\mu\nu}$-argument of
$\Gamma_k[g,\bar{g},\xi,\bar{\xi}]$ as an {\it extra background dependence}, since it is this   $\bar{g}_{\mu\nu}$-dependence 
that does not  combine with a 
corresponding  $\bar{h}_{\mu\nu}$-dependence to a full metric $ \bar{g}_{\mu\nu}+\bar{h}_{\mu\nu}\equiv g_{\mu\nu}$.
The extra $\bar{g}_{\mu\nu}$-dependence of $\Gamma_k$ has (at least) two sources, namely the one of
$\Delta_k S$ and $S_{\textrm{gf}}+S_{\textrm{gh}}$, respectively. The former disappears at $k=0$, the latter does not.

\noindent
{\bf (C)} In the construction of \cite{mr} a gauge fixing condition $\mathcal{F}_{\mu}^{\alpha\beta}[\bar{g}]h_{\alpha\beta}=0$ which is
invariant under the so called background gauge transformations \cite{back} has been employed. As a result, $\Gamma_k$ is
invariant under a simultaneous general coordinate transformation of all its arguments, including $\bar{g}_{\mu\nu}$:
\begin{equation}\label{1.6}
\Gamma_k[\Phi+\mathcal{L}_v\Phi]=\Gamma_k[\Phi],\qquad\Phi\equiv\{g_{\mu\nu},\bar{g}_{\mu\nu},\xi^{\mu},\bar{\xi}_{\mu}\}
\end{equation}
Here $\mathcal{L}_v$ denotes the Lie derivative with respect to a generating vector field $v$. At $k=0$ the standard
discussion of the background gauge technique applies \cite{back}. Hence the functional $\Gamma_{k=0}$ restricted to 
$g=\bar{g}$ or $\bar{h}=0$ is sufficient to generate all on-shell graviton Green's functions. They can be obtained
in a comparatively simply way by {\it first} setting $\bar{h}=0$ and {\it afterwards} differentiating 
$\Gamma_0[0,\xi,\bar{\xi};\bar{g}]$ with respect to $\bar{g}$.

\noindent
{\bf (D)} For vanishing ghosts\footnote{In the general case there are two more equations similar to \eqref{1.7}, involving derivatives
w.r.t. $\xi$ and $\bar{\xi}$. It follows from ghost number neutrality that those latter equations always admit the solution 
$\xi=\bar{\xi}=0$.}, the EAA implies an effective Einstein equation whose ($k$-dependent!) solution $g=g^{\textrm{sol}}$ determines
the expectation value of the metric as a functional of  $\bar{g}$:
\begin{equation}\label{1.7}
\frac{\delta}{\delta g_{\mu\nu}(x)}\widetilde{\Gamma}_k[g,\bar{g},0,0]\vert_{g=g^{\textrm{sol}}[\bar{g}]}=0
\end{equation}
Note that \eqref{1.7} involves $\widetilde{\Gamma}_k\equiv\Gamma_k +\Delta_k S$, not $\Gamma_k$ itself. In  the matter field
interpretation we may regard \eqref{1.7} as an equation for $\bar{h}^{\textrm{sol}}[\bar{g}]\equiv g^{\textrm{sol}}[\bar{g}]-\bar{g}$.
We call $\bar{g}$ a {\it selfconsistent background}, $\bar{g}=\bar{g}^{{\rm{selfcon}}}$, if it gives
rise to a vanishing fluctuation average, that is, if there are no quantum corrections to the
background metric:
\begin{equation}\label{1.8}
\bar{h}^{\textrm{sol}}[\bar{g}^{\rm{selfcon}}]=0 \quad\iff\quad g^{\textrm{sol}}=\bar{g}^{\rm{selfcon}}
\end{equation}
As $\Delta_k S[\bar{h};\bar{g}]$ is bilinear in $\bar{h}$, the condition for a selfconsistent background can be written directly in terms 
of $\Gamma_k$ itself:
\begin{equation}\label{1.9}
\frac{\delta}{\delta \bar{h}_{\mu\nu}(x)}{\Gamma}_k[\bar{h},0,0;\bar{g}^{\rm{selfcon}}]\vert_{\bar{h}=0}=0
\end{equation}
Note that in eq.\eqref{1.9} we may insert $\bar{h}=0$ only {\it after}  the functional differentiation. In order to set up the ``tadpole equation"
\eqref{1.9} we need 
to know $\Gamma_k$ at least to first order in $\bar{h}$ in an expansion about $\bar{g}$.

\noindent
{\bf (E)} On the theory space spanned by functionals of the type $\Gamma_k[g,\bar{g},\xi,\bar{\xi}]$, restricted by the condition \eqref{1.6},
the gravitational average action satisfies an exact FRGE. This FRGE was derived in \cite{mr} and up to now  all applications of the EAA 
concept in gravity  focused on finding approximate solutions to this equation \cite{mr}-\cite{je1} and to explore their physics contents
\cite{bh}-\cite{mof}. In particular the EAA-based investigations of the Asymptotic Safety scenario used this equation. For our 
present purposes a second type of exact functional equation is equally relevant to which we turn next. 

\noindent
{\bf (F)} The EAA satisfies an exact functional BRS Ward identity. To formulate it, one has to enlarge the theory space to functionals  
of the form  $\Gamma_k[g,\bar{g},\xi,\bar{\xi};\beta,\tau]$ where $\beta^{\mu\nu}$ and $\tau_{\mu}$ are sources of the BRS variation of
$\gamma_{\mu\nu}$ and $C^{\mu}$, respectively. Then, abbreviating ${\Gamma'}_k\equiv \Gamma_k-S_{gf}$, the path integral-based
definition of $\Gamma_k$ implies
\begin{equation}\label{1.20}
\int{{\rm{d}}}^dx\, \frac{1}{\sqrt{\bar{g}}}\Bigg( \frac{\delta{\Gamma'}_k }{\delta \bar{h}_{\mu\nu}}\frac{\delta{\Gamma'}_k }{\delta \beta^{\mu\nu}}
 +\frac{\delta{\Gamma'}_k }{\delta \xi^{\mu}} \frac{\delta{\Gamma'}_k }{\delta \tau_{\mu}}\Bigg)=\mathrm{Y}_k[\Gamma_k]
\end{equation}
with the trace functional 
\begin{align}\label{1.21}
\mathrm{Y}_k[\Gamma_k]= & \kappa^2 \;{\rm{Tr}}\Big[ \Big(\mathcal{R}^{\rm{grav}}_k\Big)^{\mu\nu\rho\sigma}
\Big(\Gamma^{(2)}_k+\widehat{R}_k\Big)_{\bar{h}_{\rho\sigma}\varphi}^{-1}\;
\frac{\delta^2\Gamma_k}{\sqrt{\bar{g}}\delta\varphi\;\sqrt{\bar{g}}\delta\beta^{\mu\nu}}
\Big]\nonumber\\
{} -& \sqrt{2}\;{\rm{Tr}}\Big[\mathcal{R}^{\rm{gh}}_k\Big(\Gamma^{(2)}_k+\widehat{R}_k\Big)_{\xi^{\mu}\varphi}^{-1}\;
\frac{\delta^2\Gamma_k}{\sqrt{\bar{g}}\delta\varphi\;\sqrt{\bar{g}}\delta\tau_{\mu}}
\Big]\nonumber\\
{} +& 2\alpha^{-1}\kappa^2\; {\rm{Tr}}\Big[\mathcal{R}^{\rm{gh}}_k\;\mathcal{F}_{\mu}^{\rho\sigma}\;
\Big(\Gamma^{(2)}_k+\widehat{R}_k\Big)_{\bar{h}_{\rho\sigma}\bar{\xi}_{\mu}}^{-1}
\Big]
\end{align}
where $\Gamma^{(2)}_k$ denotes the Hessian of $\Gamma_k$, and $\varphi\in\{\bar{h},\xi,\bar{\xi} \}$ is summed over. Furthermore,
$\mathcal{R}^{\rm{grav}}_k$ and $\mathcal{R}^{\rm{gh}}_k$ are the coarse graining kernels for the graviton and the ghosts, and
$\widehat{R}_k$ denotes their direct sum \cite{mr}. The standard BRS Ward identities have the structure of \eqref{1.20} with
$\mathrm{Y}_k\rightarrow 0$. The nonzero contributions to $\mathrm{Y}_k$ stem from the cutoff term $\Delta_k S$ which is not
BRS invariant. Since $\mathcal{R}_k$ vanishes for $k\rightarrow 0$ it follows that $\lim_{k\rightarrow 0}\mathrm{Y}_k=0$ so that
$\lim_{k\rightarrow 0}\Gamma_k\equiv \Gamma$ is BRS invariant in the usual way.

\noindent
{\bf (G)} The dependence of $\Gamma_k$ on the background metric $\bar{g}_{\mu\nu}$ is governed by a similar exact functional
equation\footnote{For the derivation of an analogous relation in Yang-Mills theory see Appendix A of \cite{mrcwqcd}.}:
\begin{equation}\label{1.22}
\frac{\delta}{\delta\bar{g}_{\mu\nu}(x)}\Gamma_k[g,\bar{g},\xi,\bar{\xi}]=\widetilde{\mathrm{Y}_k}[\Gamma_k]^{\mu\nu}(x)
\end{equation}
Obviously \eqref{1.22} measures the degree $\Gamma_k$ possesses an ``extra" background dependence. The functional
$\widetilde{\mathrm{Y}_k}$ on the RHS is similar to \eqref{1.21}; it consists of various traces involving $\Gamma_k$ itself.
The action appearing under the path integral, $S+S_{gf}+S_{gh}+\Delta_k S$, contains various sources of contributions
to $\mathrm{Y}_k$, in particular the extra background dependences of $S_{gf}+S_{gh}$ and $\Delta_k S$, respectively.
The former is nonzero even for $k\rightarrow 0$, the latter vanishes in this limit. 

Indeed, all coarse graining kernels have the structure
\begin{equation}\label{1.23}
\mathcal{R}_k \propto k^2R^{(0)}\Big(-\bar{D}^2/k^2\Big)
\end{equation}
where $R^{(0)}$ interpolates between zero and unity for  large and small arguments, respectively. Therefore $\mathcal{R}_k$
vanishes for $k\rightarrow 0$, and as a consequence $\Delta_k S\propto \int\, h\mathcal{R}_k h $ no longer provides
an extra background dependence. However, a crucial observation, and in fact one of the motivations for the present work, is that
the contributions stemming from $\Delta_k S$ are likely to become large in the UV limit $k\rightarrow \infty$. After all, 
$\mathcal{R}_k $ itself behaves like a divergent mass term $\propto k^2$ in this limit.

\noindent
{\bf (H)} Exact solutions to the FRGE automatically satisfy the BRS Ward identity and the $\delta/\delta\bar{g}$-equation \eqref{1.22}.
For approximate solutions to the flow equation  this is not necessarily the case. One can then evaluate the Ward identity and/or
the $\delta/\delta\bar{g}$-equation for the approximate RG trajectory and check how well these relations are satisfied. In principle
this is a useful tool in order to judge the reliability of approximations, truncations of theory space in particular.
Because of the extreme complexity of these equations this has not been done so far for gravity. In the present paper we shall use a 
simplified version of \eqref{1.22} for this purpose, however.

\subsection{QEG on truncated theory spaces}\label{subsection1.5}

The FRGE of the gravitational average action has been used in many investigations of the nonperturbative RG flow of Quantum Einstein Gravity
(QEG), in particular in the context of the Asymptotic Safety conjecture. In all of those investigations\cite{mr}-\cite{je1}
the RG flow had been projected onto a truncated theory space which can be described by the ansatz
\begin{align}\label{1.24}
\Gamma_k[g,\bar{g},\xi,\bar{\xi};\beta,\tau]= & \overline{\Gamma}_k[g]+\widehat{\Gamma}_k[g,\bar{g}]+S_{gf}[g-\bar{g};\bar{g}]+
S_{gh}[g-\bar{g},\xi,\bar{\xi};\bar{g} ]\nonumber\\
{} & -\int{\rm{d}}^dx\,\sqrt{\bar{g}}\Big( \beta^{\mu\nu}\;\mathcal{L}_{\xi}g_{\mu\nu}+\tau_{\mu}\,\xi^{\nu}\partial_{\nu}\xi^{\mu}  \Big)
\end{align}
In this ansatz the {\it classical} gauge fixing and ghost terms were pulled out of $\Gamma_k$, and also the coupling to the BRS variations 
is taken to have the same form as in the bare action. The remaining functional depends on $g_{\mu\nu}$ and $\bar{g}_{\mu\nu}$. It is further 
decomposed as $\overline{\Gamma}_k+\widehat{\Gamma}_k$ where $\overline{\Gamma}_k$ is defined by putting 
$g_{\mu\nu}$ and $\bar{g}_{\mu\nu}$ equal,
\begin{equation}\label{1.25}
\overline{\Gamma}_k[g]\equiv \Gamma_k[g,g,0,0;0,0],
\end{equation}
and $\widehat{\Gamma}_k$ is the remainder. Hence, by definition, it vanishes when the metrics are equal: $\widehat{\Gamma}_k[g,g]=0$.
Furthermore, it was argued \cite{mr} that setting  $\widehat{\Gamma}_k\equiv 0$ should be a good first approximation,
and in fact in  all calculations performed so far $\widehat{\Gamma}_k$ has been neglected essentially\footnote{At most the
effect of the running $h_{\mu\nu}$-wave function normalization ${Z}_{Nk}$ on the gauge fixing term has been taken into account
\cite{oliver1,oliver2}. This amounts to setting $\widehat{\Gamma}_k[g,\bar{g}]\propto ({Z}_{Nk}-1)S_{gf}[g-\bar{g};\bar{g}]$,
where ${Z}_{Nk}$ is given by the running of Newton's constant.}. Then, what remains to be determined from the FRGE is the 
$k$-dependence of  $\overline{\Gamma}_k[g]$, a functional of one metric variable only.

This brings us to a subtle, but important issue which will be the main topic of the present paper.

The general situation we are confronted with can be described as follows. We are given an exact RG equation on the
full theory space consisting of ``all" action functionals of a given (symmetry, etc.) type.
Somewhat symbolically, the FRGE has the structure $k\partial_k\Gamma_k=\mathcal{T}$, where $\mathcal{T}$ encodes the beta functions
of all  running couplings. These beta functions are the components of a vector field which the FRGE defines on theory
space, and the corresponding integral curves are the RG trajectories. 

Now we make a truncation ansatz for the EAA which specifies a certain
subspace of this full theory space. The idea is to study an RG flow on the subspace which is induced by the flow (vector field)
on the full space. The problem is that in general the vector field on the full space will not be tangent to the subspace and
hence it does not give rise to a flow on the latter. Stated in more practical terms, when we insert an
action from the subspace into the RHS of the FRGE, the calculation of the functional traces will produce terms different from those
present in the truncation ansatz: the RG trajectories try to ``leave" the subspace.

In order to obtain RG trajectories we must invoke a kind of generalized projection which maps the full vector field, restricted
to the subspace, onto a vector field {\it tangent to the subspace}. The result of merely restricting the full vector field  to the
subspace will not be tangent to it; it has normal components corresponding to terms in the actions we would like to discard.
Therefore the specification of a truncation involves two items: a truncation ansatz for the action, to define the subspace,
and a description for mapping the full vector field on the subspace onto a new one tangent to it.
Clearly a truncation approximates the exact flow the better the smaller the normal components of the vector field are.
Ideally one would like to find a subspace such that at least in some domain it is automatically tangent to it, without
any projection.

Now let us return to the truncation ansatz \eqref{1.24} and discuss the related projection. If we insert
the $\Gamma_k$ of \eqref{1.24}, with a possibly nontrivial $\widehat{\Gamma}_k$, into the exact form of the FRGE\footnote{
See eq.(2.32) of ref.\cite{mr}.}, we obtain a simpler flow equation for the functional
\begin{equation}\label{1.26}
\Gamma_k[g,\bar{g}]\equiv  \overline{\Gamma}_k[g]+\widehat{\Gamma}_k[g,\bar{g}]+S_{gf}[g-\bar{g};\bar{g}]
\end{equation}
It reads
\begin{equation}\label{1.27}
k\partial_k\;\Gamma_k[g,\bar{g}]=\mathcal{T}[g,\bar{g}]
\end{equation}
where
\begin{align}\label{1.28}
\mathcal{T}[g,\bar{g}]=& \;\frac{1}{2}\;{\rm{Tr}}\Big[\Big(\Gamma_k^{(2)}[g,\bar{g}]/\kappa^2+\mathcal{R}^{\rm{grav}}_k[\bar{g}]\Big)^{-1}
k\partial_k\mathcal{R}^{\rm{grav}}_k[\bar{g}]\Big]\nonumber\\
- & {\rm{Tr}}\Big[\Big(-\mathcal{M}[g,\bar{g}]+\mathcal{R}^{\rm{gh}}_k[\bar{g}]\Big)^{-1}
k\partial_k\mathcal{R}^{\rm{gh}}_k[\bar{g}]\Big]
\end{align}
Here $\Gamma_k^{(2)}$ denotes the Hessian of $\Gamma_k[g,\bar{g}]$ with respect to $g_{\mu\nu}$ at fixed $\bar{g}_{\mu\nu}$, and
$\mathcal{M}[g,\bar{g}]$ is the Faddeev-Popov kinetic operator.\footnote{See eq.(2.11) of ref.\cite{mr}.}

Let us first assume the subspace consists of  {\it all} actions of the type \eqref{1.24}, that is, $\widehat{\Gamma}_k[g,\bar{g}]$
is an arbitrary functional of two metrics, vanishing at $g=\bar{g}$. Then the vector field $\mathcal{T}[g,\bar{g}]$ in \eqref{1.28}
happens to be tangent to the subspace. In fact, the RHS of \eqref{1.27} is an arbitrary, diffeomorphically invariant functional
of $g$ and $\bar{g}$, as is its LHS. Since $S_{gf}$ is $k$-independent, the RG equation \eqref{1.27} with \eqref{1.26} reads
\begin{equation}\label{1.29}
k\partial_k\;\overline{\Gamma}_k[{g}]+k\partial_k\;\widehat{\Gamma}_k[g,\bar{g}]=\mathcal{T}[g,\bar{g}]
\end{equation}
Setting $g=\bar{g}$ in \eqref{1.29} we obtain an equation for $\overline{\Gamma}_k$ alone, and upon subtracting it from
\eqref{1.29} we get a flow equation for $\widehat{\Gamma}_k$:
\begin{align}
k\partial_k\;\overline{\Gamma}_k[{g}] &=\mathcal{T}[g,{g}]\label{1.30}\\
k\partial_k\;\widehat{\Gamma}_k[g,\bar{g}] &=\mathcal{T}[g,\bar{g}]-\mathcal{T}[g,{g}]\label{1.31}
\end{align}
Given appropriate initial conditions, the eqs.\eqref{1.30} and \eqref{1.31} suffice to determine the scale dependence of
$\overline{\Gamma}_k$ and $\widehat{\Gamma}_k$, i.e. of  all actions of the type \eqref{1.24}.

We refer to truncations of this type, involving a running functional of two metrics, $g_{\mu\nu}$  and $\bar{g}_{\mu\nu}$,
as {\it bimetric truncations}.

All truncations worked out so far in the literature \cite{mr}-\cite{frankmach} are {\it single metric truncations.}
They set $\widehat{\Gamma}_k\equiv 0$ in the general ansatz \eqref{1.24}, hence discard the second flow equation \eqref{1.31},
and set $\widehat{\Gamma}_k^{(2)}\equiv 0$ on the RHS of the first one, eq.\eqref{1.30}. In this way, the latter assumes the form,
symbolically,
\begin{equation}\label{1.32}
k\partial_k\;\overline{\Gamma}_k[{g}] =\mathcal{T}[g,\bar{g}]\vert_{\bar{g}=g,\widehat{\Gamma}_k^{(2)}= 0}
\end{equation}
This is a closed equation for $\overline{\Gamma}_{k}$, or stated differently, the vector field it defines is tangent to the $\widehat{\Gamma}_k= 0$ 
subspace.

What makes the  $\widehat{\Gamma}_k= 0$ truncations potentially dangerous is that they cannot discriminate background field monomials
in the action, such as $\int\sqrt{\bar{g}}R(\bar{g})$, from similar ones containing the dynamical metric,  $\int\sqrt{{g}}R({g})$, say.
 Hence the RG running of the  $\int\sqrt{\bar{g}}R(\bar{g})$ coefficient is combined with that of $\int\sqrt{{g}}R({g})$ into a single beta function.

The $\widehat{\Gamma}_k= 0$ truncation should provide a good approximation if the flow on the larger space is approximately
tangent to the smaller subspace defined by the additional constraint $\widehat{\Gamma}_k= 0$. This is the case if the RHS of \eqref{1.31}
is small so that the `` $\widehat{\Gamma}_k$-directions" in theory space do not get ``turned on":
$\mathcal{T}[g,\bar{g}]\approx \mathcal{T}[g,{g}] $.
This  condition is met precisely if the {\it extra} background dependence is small.

Note, however, that in
setting  $\bar{g}=g$  on the RHS of \eqref{1.31} we replace $\mathcal{R}_k[\bar{g}]$ with $\mathcal{R}_k[g]$ under the functional
traces, and one might wonder about the impact this has on the beta functions of pure $g_{\mu\nu}$  monomials, for instance.
This is in fact the topic of the present paper.

\subsection{Aim of the present paper}\label{subsection1.6}

The above discussion suggests that the degree of reliability of the $\widehat{\Gamma}_k= 0$ class of truncations  is intimately 
related to the extra background dependence of the EAA which in turn is at the very heart of the bimetric solution 
to the ``background independence" problem.

In this paper we are therefore going to analyze a first bimetric truncation, and we asses how stable the predictions of the corresponding
$\widehat{\Gamma}_k= 0$  approximation are under this generalization of the truncation. 
Indeed, we shall focus on the
non-trivial UV fixed point that is known to exist in all $\widehat{\Gamma}_k= 0$  truncations investigated so far.
The $k\rightarrow\infty$  regime is particularly susceptible to ``$\bar{g}_{\mu\nu}$
contaminations" since $\mathcal{R}_k$ diverges for
$k\rightarrow\infty$  and could possibly give rise to a large extra background dependence therefore.

For this reason  it is even the  more gratifying that we shall find a non-Gaussian fixed point (NGFP) in the RG flow of the bimetric truncation, too.
However, our analysis will not be performed within full fledged Quantum Einstein Gravity but rather a toy model which shares many features
with full QEG, in particular the existence of a NGFP  in the $\widehat{\Gamma}_k= 0$  truncation.
This toy model is the ``conformally reduced gravity" studied in \cite{creh1} and \cite{creh2}, a caricature of QEG in which the 
conformal factor of the metric is quantized in its own right, rather than the real metric degrees of freedom. In \cite{creh1} and 
\cite{creh2} a number of conceptual issues related to the Asymptotic Safety program, in particular on the role of 
``background independence", has been investigated within this comparatively simple theoretical laboratory. A  generalization
of the model including higher derivatives has been considered in \cite{crehroberto}.

For the time being the corresponding bimetric analysis of the Asymptotic Safety program within  full QEG is beyond the technical
state of art. It would require the evaluation of $\mathcal{T}[g,\bar{g}]$ for $g_{\mu\nu}$ kept different from $\bar{g}_{\mu\nu}$, by
a derivative expansion, say. We would have to calculate traces of the form ${\rm{Tr}}[f(\bar{D}^2,D^2)]$ where $f$ is a function
of two different, in general noncommuting  covariant Laplacians which involve $\bar{g}_{\mu\nu}$, and 
$g_{\mu\nu}$, respectively. There exist no standard heat kernel techniques that could be applied here.

The remaining sections of this paper are organized as follows.
In Section 2 we briefly review the discussion of conformally reduced gravity and extend it in various directions.
Using this model as our main theoretical laboratory we shall then, in Section 3, introduce bimetric truncations and
 obtain the corresponding RG flow. Section 4 is devoted to a detailed discussion of this flow and of  the general lessons it teaches us about 
full fledged QEG. We summarize our main results in Section 5.

Several discussions of a more technical nature are relegated to three appendices. Appendix A is dedicated to the evaluation of various beta
functions, in Appendix B we use the exact $\delta \Gamma_k/\delta\bar{g}_{\mu\nu}$-equation \eqref{1.22} in order to
test the quality of the truncations used, and in Appendix C we describe the relation between the effective and the bare 
fixed  point action in the bimetric setting.


\section{Conformally reduced gravity as a \\
theoretical laboratory}

Our toy model is inspired by the observation that the (Euclidean) Einstein-Hilbert action,
\begin{equation}\label{3.10}
S_{EH}[g_{\mu\nu}]=-\frac{1}{16\pi G} \int\,\textrm{d}^4x\;\sqrt{g}(R(g)-2\Lambda),
\end{equation}
when evaluated for metrics $g_{\mu\nu}=\phi^2\;\widehat{g}_{\mu\nu}$, assumes the form of a standard $\phi^4$ action:
\begin{equation}\label{3.11}
S_{EH}[\phi]=-\frac{3}{4\pi G} \int\textrm{d}^4x\sqrt{\widehat{g}}\Big(\frac{1}{2}\; \widehat{g}^{\mu\nu}\;
\partial_{\mu}\phi\;\partial_{\nu}\phi +\frac{1}{12}\widehat{R}\;\phi^2 -\frac{1}{6}\Lambda\;\phi^{4}
 \Big)
\end{equation}
Here $\widehat{g}_{\mu\nu}$ is a ``reference metric" which is fixed once and for all; in the following we usually assume it flat, whence
$\widehat{R}\equiv R(\widehat{g})=0$.  The corresponding classical equation of motion reads then
\begin{equation}\label{3.111}
\widehat{\square}\phi+\frac{2}{3}\;\Lambda\;\phi^3=0
\end{equation}
In \cite{creh1,creh2} the scalar-like theory defined by \eqref{3.11} was considered in its own right, 
detached from the original quantum field theory of metrics, and the FRGE approach has been used to quantize it.
As compared to a conventional scalar theory crucial differences arise since the background value of $\phi$ itself, denoted
$\chi_B$, determines the {\it proper} cutoff momentum a given value of $k$ corresponds to.

\subsection{The EAA setting for the toy model}\label{subsection2.1}

Let us explain the quantization and the FRGE of the toy model from a more general perspective. We start from a formal path
integral
\begin{equation}
\int\,\mathcal{D}\chi\; \exp{(-S[\chi])}
\end{equation}
representing the partition function of the scalar $\chi(x)$ with a bare action $S$, not necessarily related to $S_{EH}$ of
\eqref{3.11}. The ``microscopic" conformal factor $\chi$, the analogue of $\gamma_{\mu\nu}\equiv \bar{g}_{\mu\nu}+h_{\mu\nu}$ in
full gravity, is decomposed as $\chi(x)=\chi_B(x)+f(x)$, and the $\chi$-integral is replaced by an integral over $f$. Here 
$\chi_B(x)$ is an arbitrary but fixed background field, and $f$ the dynamical fluctuation field. The corresponding expectation values 
$\bar{f}\equiv\langle f\rangle$ and $\phi\equiv \langle\chi\rangle=\chi_B+\bar{f}$ are the counterpart of 
$\bar{h}_{\mu\nu}\equiv \langle h_{\mu\nu}\rangle$ and $g_{\mu\nu}\equiv \langle \gamma_{\mu\nu}\rangle=\bar{g}_{\mu\nu}+\bar{h}_{\mu\nu}$,
respectively.
 
By now we arrived at the path integral $\int\,\mathcal{D}f\; \exp{(-S[f+\chi_B])}$. We think of $f(x)$ as being expanded
in the eigenmodes of the covariant Laplacian $\overline{\square}$ pertaining to the background metric $\bar{g}_{\mu\nu}=\chi^2_B\,\widehat{g}_{\mu\nu}$,
whereby the measure $\mathcal{D}f$ corresponds to an integration over the expansion coefficients. The IR cutoff responsable for
the coarse graining is now implemented by introducing a smooth cutoff in the spectrum of   $\overline{\square}$, i.e. by suppressing
the contribution of all  $-\overline{\square}$ eigenmodes with eigenvalues below a given value $k^2$. In practice one replaces the path integral
by  $\int\,\mathcal{D}f\; \exp{\Big(-S[f+\chi_B]+\Delta_kS[f;\chi_B]\Big) }$ with a cutoff action which is quadratic in the fluctuation,
$$
\Delta_kS[f;\chi_B]\equiv \frac{1}{2}\int\,{\rm{d}}^4x\,\sqrt{\widehat{g}}\;f(x)\, \mathcal{R}_k[\chi_B]\, f(x),
$$
and contains a $\chi_B$-dependent integral kernel $\mathcal{R}_k[\chi_B]$.
Upon adding a source term $\int\,{\rm{d}}^4x\,\sqrt{\widehat{g}}\,Jf$ to the action the path integral equals $\exp(W_k[J;\chi_B])$ with
$W_k[J;\chi_B]$ the coarse grained generating functional of connected Green's functions. Denoting its Legendre transform by 
$\widetilde{\Gamma}[\bar{f};\chi_B]$ the definition of the effective average action for the toy model reads
\begin{equation}\label{3.12}
\Gamma_k[\bar{f};\chi_B]\equiv \widetilde{\Gamma}_k[\bar{f};\chi_B]-\Delta_kS[f;\chi_B]
\end{equation}
As in full gravity, we may alternatively regard $\Gamma_k$ as a functional of two complete metrics rather than a fluctuation and a 
background. Hence we define
\begin{equation}\label{3.13}
\Gamma_k[\phi\mathbf{,}\chi_B]\equiv \Gamma_k[\bar{f}=\phi-\chi_B\mathbf \;;\;\chi_B]
\end{equation}
Using the notation $\Gamma_k[\phi,\chi_B]$, the second argument $\chi_B$ stands for an {\it extra} background dependence, in the  sense that it
 does not appear 
combined with $\bar{f}$ as  $\chi_B+\bar{f}\equiv\phi\equiv\langle\chi\rangle$. Furthermore, in analogy with  full gravity,
we introduce the ``diagonal" functional with $\phi$ and $\chi_B$ identified,
\begin{equation}\label{3.14}
\overline{\Gamma}_k[\phi]\equiv \Gamma_k[\phi,\chi_B]\vert_{\chi_B=\phi}
\end{equation}
and the remainder $\widehat{\Gamma}_k[\phi,\chi_B]\equiv\Gamma_k[\phi,\chi_B]-\overline{\Gamma}_k[\phi]$. Thus every functional $\Gamma_k[\phi,\chi_B]$
has a unique decomposition of the form
\begin{equation}\label{3.15}
\Gamma_k[\phi,\chi_B]=\overline{\Gamma}_k[\phi]+\widehat{\Gamma}_k[\phi,\chi_B]
\end{equation}
whereby $\widehat{\Gamma}_k$ vanishes for equal fields,
\begin{equation}
\widehat{\Gamma}_k[\phi,\chi_B]\vert_{\phi=\chi_B}=0
\end{equation}

From the path integral-based definition \eqref{3.12} of the EAA its flow equation can be derived in the usual way \cite{creh1}:
\begin{equation}\label{3.20}
k\partial_k\;\Gamma_k[\bar{f};\chi_B]=\frac{1}{2}\;\textrm{Tr}\Big[ \Big( \Gamma^{(2)}_k[\bar{f};\chi_B]+ \mathcal{R}_k[\chi_B] \Big)^{-1}
k\partial_k\; \mathcal{R}_k[\chi_B] \Big ]
\end{equation}
The Hessian operator $\Gamma^{(2)}_k$ reads, in the position representation,
\begin{equation}\label{3.21}
\langle x\vert \Gamma^{(2)}_k[\bar{f};\chi_B]\vert y\rangle=\frac{1}{\sqrt{\widehat{g}(x)}\sqrt{\widehat{g}(y)}}\;
\frac{\delta^2 }{\delta\bar{f}(x)\delta\bar{f}(y)}\Gamma_k[\bar{f};\chi_B]
\end{equation}
In \cite{creh1,creh2} the coarse graining kernel $\mathcal{R}_k$ has been constructed in such a way that, when added to $\Gamma^{(2)}_k$,
it effects the replacement $-\overline{\square}\rightarrow -\overline{\square}+k^2R^{(0)}(-\overline{\square}/k^2)$, with an arbitrary shape function
$R^{(0)}$ interpolating between $R^{(0)}(0)=1$ and $R^{(0)}(\infty)=0$. For the generalized truncations considered in the present
paper this requirement is met if we choose\footnote{This choice of $\mathcal{R}_k$ generalizes the one used in earlier investigations for the
case of a position dependent $\chi_B$. If $\chi_B=const$, the operator \eqref{3.22} reduces to the one employed in \cite{creh1,creh2}.}
\begin{equation}\label{3.22}
\mathcal{R}_k[\chi_B]=\Big(-\frac{3}{4\pi G_k}\Big)\;{\chi_B}^{3}\; k^2\; R^{(0)} \Big(\frac{-\overline{\square}}{k^2}\Big)\;{\chi_B}^{-1}
\end{equation}
Note the explicit $\chi_B$-dependence of  $\mathcal{R}_k$ besides the one implicit in $\overline{\square}$, the Laplace-Beltrami operator related
to $\bar{g}_{\mu\nu}\equiv\chi^2_B\,\widehat{g}_{\mu\nu}$.

By a similar derivation one obtains the following exact equation for the extra  $\chi_B$-dependence of the EAA:
\begin{equation}\label{3.23}
\frac{\delta}{\delta\chi_B(x)}\Gamma_k[\phi,\chi_B]=\frac{1}{2}\;\textrm{Tr}\Big[ \Big( \Gamma^{(2)}_k[\phi,\chi_B]+ \mathcal{R}_k[\chi_B] \Big)^{-1}
\frac{\delta}{\delta\chi_B(x)} \mathcal{R}_k[\chi_B] \Big ]
\end{equation}
Here $ \Gamma^{(2)}_k$ involves derivatives with respect to $\phi$ at fixed $\chi_B$. Eq.\eqref{3.23} is a simplified version  of \eqref{1.22}
in full QEG.

By inserting the decomposition \eqref{3.15} into the FRGE  we obtain the following coupled system of two flow equations which is still fully equivalent
to \eqref{3.20}:
\begin{align}
k\partial_k\;\overline{\Gamma}_k[\phi] &=\mathcal{T}[\phi,\phi]\label{3.24}\\
k\partial_k\;\widehat{\Gamma}_k[\phi,\chi_B] &=\mathcal{T}[\phi,\chi_B]-\mathcal{T}[\phi,\phi]\label{3.25}
\end{align}
Here we introduced
\begin{equation}
\mathcal{T}[\phi,\chi_B]\equiv \frac{1}{2}\;\textrm{Tr}\Big[ \Big(\overline{\Gamma}_k^{(2)}[\phi]+\widehat{\Gamma}_k^{(2)}[\phi,\chi_B]
  + \mathcal{R}_k[\chi_B] \Big)^{-1}\;
k\partial_k\;\mathcal{R}_k[\chi_B] \Big ]
\end{equation}
Up to this point all equations are exact.

In the case at hand the $\widehat{\Gamma}=0$ truncations discard the second flow equation, \eqref{3.25}, and neglect the $\widehat{\Gamma}_k^{(2)}[\phi,\chi_B]$ contribution in the $\mathcal{T}[\phi,\phi]$ on the RHS of the first one. Eq.\eqref{3.24}
becomes a closed equation for $\overline{\Gamma}_k$ then:
\begin{equation}\label{3.27}
k\partial_k\;\overline{\Gamma}_k[\phi]=\frac{1}{2}\textrm{Tr}\Big[ \Big( \overline{\Gamma}^{(2)}_k[\phi]+ \mathcal{R}_k[\phi] \Big)^{-1}\;
k\partial_k\; \mathcal{R}_k[\phi] \Big ]
\end{equation}
Here we see quite explicitly why the $\widehat{\Gamma}=0$  truncations are potentially dangerous: The coarse graining kernel
under the trace, originally $\mathcal{R}_k[\chi_B]$, has now become $\mathcal{R}_k[\phi]$. Hence the cutoff terms generate
contributions to the beta functions which mix with those from the true $\phi$-terms in $\overline{\Gamma}^{(2)}_k[\phi]$!

The $\widehat{\Gamma}=0$ truncation in the toy model is completely analogous to that in Yang-Mills theory \cite{nonabavact,ym,ymrev};
in the latter case it has been successfully tested by comparison with other approaches (higher order perturbation theory, etc.).

\subsection{Examples of single-metric truncations ( $\widehat{\Gamma}=0$ )}\label{subsection2.2}

In \cite{creh1} and \cite{creh2} the FRGE of conformally reduced gravity was solved in various truncations of the 
$\widehat{\Gamma}=0$ type. The simplest one is the ``conformally reduced Einstein-Hilbert" (CREH) truncation;
here the ansatz for the EAA has exactly the structure of the classical action \eqref{3.11}, with a running Newton and cosmological
constant, though:
\begin{equation}\label{3.30}
\Gamma_k[\phi,\chi_B] =-\frac{3}{4\pi G_k}\int\,\textrm{d}^4x\;\sqrt{\widehat{g}}\;\Big\{ 
\frac{1}{2}\; \widehat{g}^{\mu\nu}\;\partial_{\mu}\phi\; \partial_{\nu}\phi
+ \frac{1}{12}\widehat{R}\; \phi^2-\frac{1}{6}\Lambda_k\;\phi^4 \Big\}
\end{equation}
In \cite{creh2} a generalization motivated by the ``local potential approximation" frequently used in standard scalar theories \cite{avactrev}
was employed; it retains the classical kinetic term of \eqref{3.30} but allows for an arbitrary potential:
\begin{equation}\label{3.31}
\Gamma_k[\phi,\chi_B] =-\frac{3}{4\pi G_k}\int\,\textrm{d}^4x\;\sqrt{\widehat{g}}\;\Big\{ 
\frac{1}{2}\; \widehat{g}^{\mu\nu}\;\partial_{\mu}\phi\; \partial_{\nu}\phi
+ F_k(\phi)\Big\}
\end{equation}

The crucial feature of the functionals \eqref{3.30} and \eqref{3.31} is that actually they have {\it no extra dependence on} $\chi_B$,
that is, the background field $\chi_B$ always appears combined with the fluctuation $\bar{f}$ to form a  complete field
$\phi=\chi_B+\bar{f}$. Hence we have $\Gamma_k[\phi,\chi_B]=\overline{\Gamma}_k[\phi]$ and $\widehat{\Gamma}[\phi,\chi_B]=0$.

In \cite{creh1} the flow corresponding to the CREH truncation \eqref{3.30} has been worked out whereby the coarse graining operator
\eqref{3.22} was used. This operator is designed in such a way that the cutoff scale $k$ is proper with respect to the background metric
$\bar{g}_{\mu\nu}=\chi^2_B\,\widehat{g}_{\mu\nu}$. It was found that, with this $\mathcal{R}_k$, the RG flow is qualitatively very similar
 to the one in full QEG; in particular both a Gaussian and a non Gaussian fixed point  were found to exist in this truncation.

If instead  $\mathcal{R}_k$ is tailored in such  a way that $k$ becomes proper with respect to the metric $\widehat{g}_{\mu\nu}$, the RG
flow is that of a conventional scalar theory, and no NGFP exists.

In \cite{creh1} it was argued that the first choice of  $\mathcal{R}_k$ is the correct one to be used in gravity since only this choice 
respects the principle of ``background independence", while the second makes use of a rigid structure, the reference metric
 $\widehat{g}_{\mu\nu}$, which does not even have an analogue in full QEG.

The flow in the local potential approximation \eqref{3.31} has been worked out in \cite{creh2} where in particular phase transitions to
a phase of gravity with unbroken diffeomorphism invariance were studied in a simple setting. For a higher derivative generalization
see \cite{crehroberto}.


\section{Bimetric truncations}

\subsection{Generalized local potential approximation}\label{subsection3.1}

In the following we employ a generalized truncation ansatz which will allow us to disentangle the $\phi$- and $\chi_B$-dependencies
of the EAA. We no longer identify the dynamical metric $g_{\mu\nu}\equiv\phi^2\,\widehat{g}_{\mu\nu}$ with the background metric 
 $\bar{g}_{\mu\nu}\equiv\chi^2_B\,\widehat{g}_{\mu\nu}$ as in \eqref{3.27}. The ansatz has a nontrivial extra $\chi_B$ dependence now. It
reads\footnote{In ref.\cite{floper}, Floreanini and Percacci have performed a  similar calculation with two independent conformal factors
 in a perturbatively renormalizable gravity model.}
\begin{equation}\label{3.40}
\Gamma_k[\phi,\chi_B] =-\frac{3}{4\pi }\int\,\textrm{d}^4x\;\sqrt{\widehat{g}}\;\Big\{ 
\frac{1}{2 G_k}\;\phi(-\widehat{\square})\phi +\frac{1}{2 G_k^B}\;\chi_B(-\widehat{\square})\chi_B
+ \frac{1}{ G_k}F_k(\phi,\chi_B)\Big\}
\end{equation}
This ansatz differs from \eqref{3.31} by a separate kinetic term for the background field and an extra $\chi_B$ dependence of the potential
$F_k(\phi,\chi_B)$. Clearly the functional $\widehat{\Gamma}_k[\phi,\chi_B]$ related to \eqref{3.40}  is nonvanishing for generic fields.
This ansatz may be regarded a generalized local potential ansatz for two ``scalars";
if we require $\Gamma_k$ to be separately invariant under $\phi\rightarrow -\phi$ and   $\chi_B\rightarrow -\chi_B$ there is no cross
term $\phi\widehat{\square}\chi_B$. 

Note that that there exist two versions of Newton's constant now: the prefactor of $\phi(-\widehat{\square})\phi$
involves the ordinary Newton constant $G_k$ associated with the (self-) couplings of the dynamical gravitational field $\phi$, while the prefactor
of $\chi_B(-\widehat{\square})\chi_B$ contains a kind of background Newton constant $G_k^B$. (In the potential term we pulled out
a factor of $G_k$ to facilitate the comparison with \eqref{3.31}.) The scale dependence of the two Newton constants is governed by their 
respective anomalous dimension, defined as
\begin{align}\label{3.41}
\eta_N \equiv k\partial_k\;\ln{G_k},&\quad \eta_N^B\equiv k\partial_k\;\ln{G_k^B}
\end{align}

In order to project out the $\chi_B(-\widehat{\square})\chi_B$ term we must allow for a $x$-dependent background field in the following; in 
\cite{creh1,creh2} a constant one had been sufficient.

In the following we assume $\widehat{g}_{\mu\nu}$ to be a flat metric on a manifold with ${R}^4$-topology. We shall set 
$\widehat{g}_{\mu\nu}=\delta_{\mu\nu}$ where convenient.

\subsection{The 3-parameter potential ansatz}\label{subsection3.2}

For an explicit solution of the differential equations we shall impose a further truncation on the ansatz \eqref{3.40}. We assume
that $F_k(\phi,\chi_B)$ involves only three running couplings, multiplying the monomials $\phi^4$, $\phi^2\chi^2_B$ and $\chi_B^4$,
respectively:
\begin{equation}\label{3.412}
F_k(\phi,\chi_B) =-\frac{\Lambda_k}{6}\;\phi^4 + \frac{M_k}{2}\;\phi^2{\chi_B}^2-\frac{1}{6}\,\frac{G_k}{G_k^B}\,\Lambda_k^B\;\chi_B^4\\
\end{equation}
This ansatz allows us to disentangle the $\phi$- and the $\chi_B$-contributions, respectively, to the cosmological constant term in
$\overline{\Gamma}_k[\phi]\equiv \Gamma_k[\phi,\chi_B=\phi]$.

The actions $\overline{\Gamma}_k[\phi]$ and $\widehat{\Gamma}_k[\phi,\chi_B]$ contain the potentials $\bar{F}_k(\phi)$ and 
$\widehat{F}_k(\phi,\chi_B)$ with  $\bar{F}_k(\phi)\equiv F_k(\phi,\phi)$ and $\widehat{F}_k(\phi,\chi_B)=F_k(\phi,\chi_B)-F_k(\phi,\phi)\neq 0$,
respectively.

The single metric potential  $\bar{F}_k(\phi)$ contains a cosmological constant term $\propto \phi^4$; this is the one
whose running has been computed in earlier studies. All three monomials in the ansatz \eqref{3.412} contribute to this
term upon equating the fields:
$$
\bar{F}_k(\phi)=F_k(\phi,\chi_B=\phi)\propto \phi^4
$$

The 3-parameter potential \eqref{3.412} is further motivated by the fact that there is a natural way of projecting the flow on the corresponding
truncation subspace (see below and Appendix A), and by its natural interpretation in the language of conventional scalar field theory. In fact, 
besides the ``true" cosmological constant $\Lambda_k$ related to the dynamical field and the ``background" one $\Lambda_k^B$, analogously
related to $\chi_B$, the potential contains the mixed term $\propto \chi^2_B\phi^2$. As we shall see the later it is closely related to a conventional 
mass term.

The functional \eqref{3.40} with the special potential \eqref{3.412} can be written in the following suggestive form:
\begin{align}\label{3.413}
\Gamma_k[\phi,\chi_B] & = -\frac{1}{16\pi G_k}\int\,\textrm{d}^4x\;\sqrt{g}\Big(R(g)-2\Lambda_k\Big)
-\frac{1}{16\pi G_k^B}\int\,\textrm{d}^4x\;\sqrt{\bar{g}}\Big(R(\bar{g})-2\Lambda_k^B\Big)\nonumber\\
{}& -\frac{3M_k}{8\pi G_k}\int\,\textrm{d}^4x\;\Big[\sqrt{\bar{g}}\sqrt{g}\Big]^{1/2}
\end{align}
Here it is understood that the RHS of this equation is evaluated for the metrics $g_{\mu\nu}=\phi^2\delta_{\mu\nu}$ and 
$\bar{g}_{\mu\nu}=\chi^2_B\delta_{\mu\nu}$. Obviously the above functional consists of two separate Einstein-Hilbert actions
for $g_{\mu\nu}$ and $\bar{g}_{\mu\nu}$, respectively, plus a novel non-derivative term which couples the two metrics. Unusual
as it looks, it is precisely the kind of terms that is expected to arise in the effective average action 
$\Gamma_k[g_{\mu\nu},\bar{g}_{\mu\nu}]$ of full quantum gravity. In fact, the RHS of \eqref{3.413} is invariant under
simultaneous diffeomorphisms of  $g_{\mu\nu}$ and $\bar{g}_{\mu\nu}$, as it should.

It is instructive to write down the tadpole equation for the toymodel. The analogue of eq.\eqref{1.9} for a selfconsistent
background $\chi_B\equiv \chi_B^{\textrm{selfcon}}$ reads, with the general truncation \eqref{3.40}
\begin{equation}\label{3.414}
\widehat{\square}\chi_B-F_k'(\chi_B,\chi_B)=0
\end{equation}
Here the prime denotes a derivative with respect to the $\phi$-argument before $\phi=\chi_B$ is set. For the 3-parameter potential
\eqref{3.412} we find, for instance,
\begin{equation}\label{3.415}
\widehat{\square}\chi_B+\frac{2}{3}\Lambda_k^{\textrm{scb}}\;\chi^3_B=0
\end{equation}
with the selfconsistent background (scb) parameter
\begin{equation}\label{3.416}
\Lambda_k^{\textrm{scb}}=\Lambda_k-\frac{3}{2}\;M_k
\end{equation}
While the equation \eqref{3.415} which governs consistent background fields has the same structure as the classical field equation
\eqref{3.111}, or the one following from the $\widehat{\Gamma}=0$ truncation \eqref{3.30}, it is neither the classical nor the 
running cosmological $\Lambda_k$ that enters here. The curvature scale of selfconsistent backgrounds is set by the combination \eqref{3.416}
involving $M_k$.

\subsection{The coarse graining kernel}

It is important to note that pure $\chi_B$-terms in $\Gamma_k$ do not contribute to the RHS of the FRGE; they vanish when we perform the
$\delta/\delta \bar{f}$-derivatives in $\Gamma_k^{(2)}[\bar{f};\chi_B]$ which amount to a $\delta/\delta \phi$ -derivative in the
$(\phi,\chi_B)$ -language. In fact from \eqref{3.40} it follows that, in operator notation:
\begin{equation}\label{3.42}
\Gamma_k^{(2)}[\phi, \chi_B]=-\frac{3}{4\pi G_k}\Big[-\widehat{\square}+  F^{''}_k(\phi, \chi_B)   \Big]
\end{equation}
Here and in the following a prime denotes a derivative w.r.t. the dynamical field, $\phi$. Obviously only mixed $\phi$-$\chi_B$ potential
terms with at least two powers of $\phi$ contribute to $\Gamma_k^{(2)}$.

As it stands, \eqref{3.42} involves the Laplacian $\widehat{\square}$ built from $\widehat{g}_{\mu\nu}$. For the construction of
$\mathcal{R}_k$ it is more convenient to express  $\Gamma_k^{(2)}$ in terms of $\overline{\square}$, the analogous Laplace-Beltrami
operator from $\bar{g}_{\mu\nu}\equiv\chi^2_B\,\widehat{g}_{\mu\nu}$. The two Laplacians are related by a well-known identity for Weyl rescalings,
$$
-\widehat{\square}+ \frac{1}{6}\widehat{R}=\chi_B^3\;\Big[ -\overline{\square}+\frac{1}{6}\bar{R}\Big]\; \chi_B^{-1}\nonumber
$$
Here $\bar{R}\equiv R(\bar{g})=R(\chi^2_B\widehat{g})$  and $\widehat{R}\equiv R(\widehat{g})$. Since, in our case, $\widehat{R}=0$, we may rewrite
\eqref{3.42} as
\begin{equation}\label{3.43}
\Gamma_k^{(2)}[\phi, \chi_B]=\chi_B^3\;\Big(-\frac{3}{4\pi G_k}\Big)\Big[-\overline{\square}+ \frac{1}{6}\bar{R}+
\frac{1}{\chi^2_B} F^{''}_k(\phi, \chi_B)\Big]\; \chi_B^{-1}
\end{equation}
Using the representation \eqref{3.43} it is easy to see that the $\mathcal{R}_k$-operator announced above in eq.\eqref{3.22} is indeed the correct
one for implementing the ``background independent" coarse graining. Combining \eqref{3.22} with \eqref{3.43} we have
\begin{equation}\label{3.44}
\Gamma_k^{(2)}+\mathcal{R}_k=\chi_B^3\;\Upsilon\; \chi_B^{-1}
\end{equation}
with
\begin{equation}\label{3.45}
\Upsilon\equiv \Big(-\frac{3}{4\pi G_k}\Big)\Big[-\overline{\square}+k^2\, R^{(0)}\Big(\frac{-\overline{\square}}{k^2}\Big)+ \frac{1}{6}\bar{R}+
\frac{1}{\chi^2_B} F^{''}_k(\phi, \chi_B)\Big]
\end{equation}
We see that, as required, adding $\mathcal{R}_k$ to $\Gamma_k^{(2)}$ replaces $-\overline{\square}$ with 
$-\overline{\square}+k^2\, R^{(0)}\Big(\frac{-\overline{\square}}{k^2}\Big)$. As a result, the $-\overline{\square}$-eigenmodes with eigenvalues $\lesssim k^2$
get suppressed  by a mass term in their inverse propagator $(-\overline{\square}+k^2+\cdots)$. As it was explained in \cite{creh1},
this is exactly what an $\mathcal{R}_k$ respecting ``background independence" must do.

\subsection{The beta functions}\label{subsection3.4}

Now we are ready to write down the flow equation on the truncation subspace. Inserting \eqref{3.44} along with
$k\partial_k \mathcal{R}_k= \chi_B^3\, \rho \, \chi_B^{-1}$ where $\rho\equiv k\partial_k\{(-3/4\pi G_k)\,k^2\,R^{(0)}(-\overline{\square}/k^2)\}$
into the exact equation \eqref{3.20} we obtain
$
k\partial_k\Gamma_k[\phi, \chi_B]=\frac{1}{2}{\rm{Tr}}\Big[\Big( \chi_B^3\,\Upsilon\, \chi_B^{-1}\Big)^{-1} \chi_B^3\, \rho \, \chi_B^{-1} \Big].
$
Even though $\Upsilon$ and $\rho$ do not commute with $\chi_B$ the cyclicity of the trace allows us to simplify the RHS of this equation:
$k\partial_k\;\Gamma_k[\phi, \chi_B]=\tfrac{1}{2}{\rm{Tr}}\big[\Upsilon^{-1}\rho \big]$. If we finally insert the $\Gamma_k$-ansatz \eqref{3.40}
on its LHS we obtain, using \eqref{3.41},
\begin{align}\label{3.46}
\frac{3}{4\pi} \int\,{\rm{d}}^4x\sqrt{\widehat{g}}&\,\Big\{ \frac{\eta_N}{2 G_k}\,\phi(-\widehat{\square})\phi+
\frac{\eta_N^B}{2 G_k^B}\,\chi_B(-\widehat{\square})\chi_B- 
\frac{1}{G_k}(k\partial_k\; - \eta_N)F_k(\phi, \chi_B)
\Big\}\nonumber\\
=&\; k^2 \;{\rm{Tr}}\Big[\Big(-\overline{\square}+k^2\, R^{(0)}\Big(\frac{-\overline{\square}}{k^2}\Big)+ \frac{1}{6}\bar{R}+
\frac{1}{\chi^2_B} F^{''}_k(\phi, \chi_B)\Big)^{-1}\nonumber\\
\times&  \Big( \big(1-\frac{\eta_N}{2}\big)R^{(0)}\Big(\frac{-\overline{\square}}{k^2}\Big)- 
 \Big(\frac{-\overline{\square}}{k^2}\Big)  {R^{(0)}}'\Big(\frac{-\overline{\square}}{k^2}\Big)\Big)\Big]
\end{align}
Eq. \eqref{3.46} is the ``master equation" for finding all beta functions of the truncated system. We have to perform a simultaneous derivative
expansion of the functional trace, retaining all non-derivative terms, as well as the terms with two derivatives and two powers of 
$\phi$ and $\chi_B$, respectively. We can then match the corresponding monomials on both sides of the FRGE and read off
the beta functions for the potential, $G_k$ and $G_k^B$, respectively. The corresponding calculations are discussed in 
Appendix A; here we display the results only.

Without restricting the form of the potential we can derive a partial differential equation which governs the $k$-dependence
of $F_k(\phi,\chi_B)$ regarded as an arbitrary function of two field variables. It reads
\begin{equation}\label{3.47}
\Big( k\partial_k -\eta_N \Big)F_k(\phi,\chi_B) =  -\frac{G_k}{24\pi}\big(1-\frac{\eta_N}{6}\big)\frac{(\chi_B k)^6 
}{(\chi_B k)^2+\partial^2_{\phi}F_k(\phi,\chi_B)}
\end{equation}

It will be convenient to rewrite \eqref{3.47} in dimensionless terms. As $\phi$ and $\chi_B$ have the dimension of a  length\footnote{We assign the 
following canonical mass dimensions: $[x^{\mu}]=0$, $[g_{\mu\nu},\bar{g}_{\mu\nu}]=-2$, $[\widehat{g}_{\mu\nu}]=0$, $[\phi,\chi_B,\bar{f}]=-1$.
The running couplings have $[G,G^B]=-2$ and $[\Lambda,\Lambda^B,M]=2$.}
the quantities
\begin{align}\label{3.48}
\varphi\equiv k\;\phi,&\quad b\equiv k\;\chi_B
\end{align}
are dimensionless and,
\begin{equation}\label{3.49}
Y_k(\varphi,b)\equiv k^2 \;F_k(\varphi/k,b/k)
\end{equation}
is a dimensionless function of two dimensionless field arguments. 
We shall also need the dimensionless coupling constants
\begin{align}\label{3.50}
g_k\equiv k^2\,G_k, \quad \lambda_k\equiv & \Lambda_k/k^2, \quad m_k\equiv M_k/k^2
\nonumber\\
 g_k^B\equiv k^2\,G_k^B, & \quad \lambda_k^B\equiv \Lambda_k^B/k^2. 
\end{align}
The flow equation for the dimensionless potential assumes the form
\begin{equation}\label{3.51}
\Big( k\partial_k\;+\varphi\partial_{\varphi}+b\partial_b -\eta_N -2 \Big)Y_k(\varphi,b) =  -\frac{g_k}{24\pi}\;
\big(1-\frac{\eta_N}{6}\big)\frac{b^6 }{b^2+\partial_{\varphi}^2 Y_k(\varphi,b)}
\end{equation}

If we truncate further and assume $F_k$ to be of the 3-parameter form \eqref{3.412} the FRGE boils down to the following system of 5 coupled
ordinary differential equations for the running couplings $\{g_k,\lambda_k,m_k;g_k^B,\lambda_k^B \}$:
\begin{subequations}\label{3.52}
\begin{align}
k\partial_k\; g_k= & \beta_g\equiv [2+\eta_N(g_k,\lambda_k,m_k)]\;g_k
\label{3.52a}\\
k\partial_k\; \lambda_k = & \beta_{\lambda}\equiv (\eta_N-2)\lambda_k+ 
\frac{g_k}{\pi}\Big(1-\frac{1}{6}\eta_N\Big)\frac{\lambda_k^2}{(1+m_k)^3}
\label{3.52b}\\
k\partial_k\; m_k = & \beta_{m}\equiv (\eta_N -2)m_k-
\frac{g_k}{6\pi}\Big(1-\frac{1}{6}\eta_N\Big)\frac{\lambda_k}{(1+m_k)^2}
\label{3.52c}\\
k\partial_k\; g_k^B= & \beta_g^B \equiv [2 +\eta_N^B(g_k,\lambda_k,m_k;g_k^B)]\;g_k^B
\label{3.52d}\\
k\partial_k\; \lambda_k^B = & \beta_{\lambda}^B\equiv (\eta_N^B-2)\lambda_k^B+ 
\frac{g_k^B}{4\pi}\Big(1-\frac{1}{6}\eta_N\Big)\frac{1}{(1+m_k)}
\label{3.52e}
\end{align}
\end{subequations}
The anomalous dimensions are
\begin{subequations}\label{3.53}
\begin{align}
\eta_N(g_k,\lambda_k,m_k) =& -\frac{2}{3\pi}\frac{g_k\lambda_k^2}{(1+m_k-2\lambda_k)^4}
\label{3.53a}\\
\eta_N^B(g_k,\lambda_k,m_k;g_k^B) =& g_k^B\,\Big[B_1(\lambda_k-\frac{1}{2}m_k)+\eta_N B_2(\lambda_k-\frac{1}{2}m_k)-\frac{\eta_N}{g_k}\Big]
\label{3.53b}
\end{align}
\end{subequations}
The functions $B_1$ and $B_2$ are defined in Appendix A. 

There are two special cases in which the structure of $\eta_N^B$ simplifies:
\begin{equation}
\eta_N^B(g_k,\lambda_k=0,m_k=0;g_k^B) = \frac{g_k^B}{12\pi}
\end{equation}
\begin{equation}
\eta_N^B(g_k=0,\lambda_k,m_k;g_k^B) = g_k^B\,\Big[ B_1(\lambda_k-\frac{1}{2}m_k)+\frac{2}{3\pi}\frac{\lambda_k^2}{(1+m_k-2\lambda_k)^4}\Big]
\end{equation}

The five equations \eqref{3.52} decouple to some extent. The first three of them, \eqref{3.52a}, \eqref{3.52b}, \eqref{3.52c}, close among themselves.
They do not involve $g_k^B$ and $\lambda_k^B$, and they are sufficient to find the $k$-dependence of $g_k$, $\lambda_k$ and $m_k$. We 
shall refer them as the $\{g,\lambda,m\}$ subsystem.

Having found some solution of this subsystem we may insert it into the remaining equations \eqref{3.52d}, \eqref{3.52e}. In this way they become a 
(non-autonomous) system of two equations for the remaining unknowns, namely $g_k^B$ and $\lambda_k^B$.

This decoupled structure of the equations facilitates in particular the search for fixed points of the RG flow. One first determinates those of the 
$\{g,\lambda,m\}$ subsystem and, in a second step, inserts their coordinates $(g_*,\lambda_*,m_*)$ into the remaining beta functions
$\beta_g^B$ and $\beta_{\lambda}^B$. If they admit a common zero $(g_*^B,\lambda_*^B)$, the fixed point of the subsystem extends to 
the fixed point of the five dimensional flow.


\section{Properties of the bimetric RG flow}

\subsection{Comparison with standard scalar field theory on a\\
 rigid background}

The gravitational average action $\Gamma_k[g_{\mu\nu},\bar{g}_{\mu\nu}]$ is a functional of two metrics, and the full information is 
available only if $\bar{g}_{\mu\nu}$ is kept arbitrary. In particular this is necessary for setting up an FRGE. Likewise, in the conformally 
reduced case, $\Gamma_k[\phi,\chi_B]$ is defined for an arbitrary $\chi_B$, in accordance with  the requirement of ``background
independence". Nevertheless,  $\Gamma_k[\phi,\chi_B]$ should contain also the information about the beta functions one computes in the 
standard rigid-background approach. 

To see this one fixes $\bar{g}_{\mu\nu}$ once and for all, say as $\bar{g}_{\mu\nu}=l^2\delta_{\mu\nu}$
with some constant length scale $l$; this amounts to setting $\chi_B=l$ everywhere. Let us consider the reduced functional
\begin{equation}\label{4.1}
\Gamma_k[\phi]^{\rm{rigid}}\equiv \Gamma_k[\phi,\chi_B=l]
\end{equation}
Within the above truncation it contains the three-parameter potential
\begin{equation}\label{4.2}
F_k(\phi)^{\rm{rigid}}\equiv F_k(\phi,l)=-\frac{1}{6}\,\Big(\frac{G_k}{G_k^B}\Big)\;\Lambda_k^B\;l^4
+\frac{1}{2}(M_k\,l^2)\;\phi^2-\frac{1}{6}\Lambda_k\;\phi^4
\end{equation}
From the ``rigid" perspective we are dealing with a conventional $\mathbb{Z}_2$-symmetric, single component scalar theory, with a running $\phi^4$-coupling
proportional to $\Lambda_k$, a $(\rm{mass})^2$ parameter $\propto M_k$, and a $\phi$-independent term $\propto\Lambda_k^B $ in its
potential. The last term is physically irrelevant from the rigid point of view. The  $\Gamma_k$ of eq.\eqref{3.40} has also
a nontrivial wavefunction normalization $\propto 1/G_k$. Up to a minus sign to which we return in a moment, the latter can be removed
by scaling the field. Eq.\eqref{3.11} with $\widehat{g}_{\mu\nu}=\delta_{\mu\nu}$ implies
\begin{equation}\label{4.3}
S[\sqrt{4\pi G/3}\,\phi]=  \int\,{\rm{d}}^4x\Big\{    -\frac{1}{2}\delta^{\mu\nu}\partial_{\mu}\phi\partial_{\nu}\phi 
+\frac{\mathnormal{u}}{4!}\;\phi^4\Big\}
\end{equation}
This makes it clear that the combination of couplings
\begin{equation}\label{4.4}
u_k\equiv\frac{16\pi}{3}G_k\Lambda_k=\frac{16\pi}{3}g_k\lambda_k
\end{equation}
plays the role of the conventionally normalized $\phi^4$-coupling constant in the rigid scheme. 

Taking advantage of the beta 
functions $\beta_g$ and $\beta_{\lambda}$ from eqs.\eqref{3.52a} and \eqref{3.52b} we obtain the following RG equation for $u_k$:
\begin{equation}\label{4.5}
k\partial_k\; u_k =\beta_u\equiv  2\eta_N u_k+\frac{3}{16\pi^2}\; \big(1-\frac{\eta_N}{6}\big)\;\frac{u_k^2}{(1+m_k)^3}
\end{equation}
This is precisely what one would also obtain from the truncated FRGE in a conventional scalar calculation \cite{avactrev}. The factor
$(1-\eta_N/6)$ in the last term is an automatic ``RG improvement" due to the wave function renormalization (here $\propto 1/G_k$)
present in $\mathcal{R}_k$, and the factor $1/(1+m_k)^3$ originates from a threshold function; it describes the decoupling at large masses,
$m_k\gg 1$, i.e.   $M_k\gg k^2$. Omitting these refinements we are left with the purely perturbative beta function
$\beta_u^{\rm{pert}}=2\eta_N u+\tfrac{3}{16\pi^2}u^2$. Note that this function can have a zero, corresponding to a fixed point
of the  $u$-evolution  only if the product $ \eta_N u$ can become negative.

As for the possibility of a negative value of $ \eta_N u$ there is a crucial difference between a standard scalar and the gravity
toy model: the ''wrong" negative sign of the kinetic term in \eqref{4.3}. Even though this is the minus sign behind the notorious conformal factor 
instability\footnote{We expect that in full quantum gravity, described by a sufficiently general truncation, the instability is
cured by higher derivative terms which  contribute positively to the Euclidean action. This scenario is known to be realized in the
$R+R^2$ truncation of QEG \cite{oliver2} where the inverse propagator is of the type $-p^2+p^4$. (See also ref.\cite{oliverunstable}.)},
it has a deep physical meaning which shows that it {\it must} occur at this place if the toy model is supposed to mimic gravity:
It simply encodes the fact that gravity is a universally {\it attractive} theory, in the sense that like ``charges" (masses, etc.),
and there are no others, attract rather than repel one another\footnote{This is most easily seen by comparing  the Newtonian 
approximation of classical General Relativity to Electrostatics. In the former case the field equation is 
$+\nabla^2\varphi_{\rm{grav}}=4\pi G\rho_{\rm{grav}}$, while in the latter $-\nabla^2\varphi_{\rm{el}}=4\pi\rho_{\rm{el}}$.
The first Poisson equation obtains from the {\it unstable} Einstein-Hilbert action in the Newton limit ($\varphi_{\rm{grav}}$ is
closely related to $\phi$), the second from the {\it stable} Maxwell action. Their relative minus sign is obviously in one-to-one
correspondence to the wrong sign of the kinetic term of $\varphi_{\rm{grav}}$ or $\phi$. }.

In the conventional, stable, scalar theory the anomalous dimension is either zero, in the symmetric phase, or positive, in the broken
phase. The corresponding $\eta_N$ for the gravity toy model is given by eq.\eqref{3.53a}. If $g > 0$ it is always
{\it negative}, and this is a consequence of the wrong sign of the kinetic term. In the regime $m,\lambda\ll 1$ for instance, one has
$\eta_N=-\tfrac{2}{3\pi}\,g\,\lambda^2$. We shall see that thanks to this negative anomalous dimension the gravity
model is asymptotically safe, while the conventional stable theory is not.

The calculation of $\eta_N$ in the gravity model is similar to that in the broken phase of standard scalar theory. The anomalous
dimension of the latter is proportional to the (squared) vacuum expectation value of the field, $v$. In the corresponding calculation
 of $\eta_N$ in the gravity model the role of $v$ is played by the constant value of $\chi_B$ which is necessarily nonzero.

\subsection{How the vacuum energy relates to the\\
 cosmological constant(s)} 

We continue the analysis of the RG equations by specializing for the regime where $\eta_N,\eta_N^B\ll1$ and $m_k\ll1$ which corresponds
to lowest order perturbation theory. Here the dimensionful Newton constants $G_k$ and $G_k^B$ do not run; we shall denote their
constant values as $\bar{G}$ and $\bar{G}^B$, respectively. For the running of the other three dimensionful couplings the equations
\eqref{3.52} imply, in this approximation,
\begin{subequations}\label{4.6}
\begin{align}
k\partial_k\;\Lambda_k= &\frac{1}{\pi}\;\bar{G}\;\Lambda_k^2
\label{4.6a}\\
k\partial_k\;M_k=&-\frac{1}{6\pi}\;\bar{G}\;\Lambda_k\, k^2
\label{4.6b}\\
k\partial_k\;\Lambda_k^B=&\frac{1}{4\pi}\;\bar{G}^B\, k^4
\label{4.6c}
\end{align}
\end{subequations}

The first equation \eqref{4.6a} for the genuine cosmological constant $\Lambda_k$ is easily solved; we find that it runs only logarithmically:
\begin{equation}\label{4.7}
\Lambda_k\propto 1/\ln{k}
\end{equation}
This logarithmic running is exactly what one would expect from the rigid perspective: Being the prefactor of the $\phi^4$-monomial,
the genuine cosmological constant must display the well known logarithmic running of a standard $\phi^4$ coupling in perturbation 
theory.

Since the $k$-dependence of $\Lambda_k$ is very weak we may solve \eqref{4.6b} by approximating $\Lambda_k=\rm{const}$ on its
RHS. Thus, up to logarithmic corrections,
\begin{equation}\label{4.8}
M_k= -\frac{1}{12\pi}\bar{G}\Lambda\, k^2 +\rm{const}
\end{equation}
Again, from  the rigid point of view this is the expected result: $M_k$, the prefactor of $\phi^2$, plays the role of the mass
square which is well known to renormalize quadratically.

The equation \eqref{4.6c} for the background cosmological constant $\Lambda_k^B$ is decoupled from the other two; it
integrates to
\begin{equation}\label{4.9}
\Lambda_k^B= \frac{1}{16\pi}\bar{G}^B\, k^4 +\rm{const}
\end{equation}
The running parameter $\Lambda_k^B$ is physically irrelevant in the rigid theory, it enters a field independent term only,
but it is crucial for a proper understanding of the gravity theory.

In fact, this last result is rather striking and important conceptually. Beginning with Pauli \cite{straumann}
many physicists argued that every mode of a quantized matter field in its ground state should contribute the zero point energy
$\tfrac{1}{2}\hbar\omega$ to the energy density of the vacuum. It was furthermore argued that this contribution 
to the vacuum energy should be part of the cosmological constant, and as such it should contribute to the curvature of the
Universe. If one sums up the zero point energies of a massless free  field from $\omega=0$ to an UV cutoff
$\omega_{\rm{max}}$ one finds an energy density that grows proportional to  $ \omega_{\rm{max}}^4$. As is well known
\cite{coscon,straumann} even moderate values of $\omega_{\rm{max}}$ lead to tremendously large contributions to the cosmological 
constant, exceeding the observed value by many orders of magnitude. This is (part of) the notorious cosmological constant
problem.

The average action ``knows" about this sum of the zero point energies. It corresponds to a one-loop contribution and can be made
manifest in the corresponding approximation to $\Gamma_k$; it gives rise to  a characteristic $k^4$-dependence then.\footnote{
In this respect the gravitational field behaves in exactly the same way as a massless matter field.}

Moreover it is well known \cite{mr} that {\it in a single metric truncation }of full QEG ,  to lowest order in $G_k$,
the (single) cosmological constant runs proportional to $k^4$. In the above calculation we used on refined truncation
which can distinguish contributions to $\Lambda_k\int\sqrt{g}$ and $\Lambda_k^B\int\sqrt{\bar{g}}$. The results \eqref{4.7}
and \eqref{4.9} show {\it the background cosmological constant $\Lambda_k^B$, rather than the genuine one, $\Lambda_k$,
displays the fast $k^4-$running due to the summed zero point energies. } The genuine cosmological constant, the coefficient
of the volume element provided the dynamical metric, $\sqrt{g}$, has only a very weak, logarithmic scale dependence.

Clearly, this observation is  relevant to the cosmological constant problem:
Only $\Lambda_k$, but not $\Lambda_k^B$, occurs in the effective Einstein equations \eqref{1.7} or \eqref{1.9};
it is the genuine rather than the background cosmological constant that determines  the average curvature of spacetime.
 But contrary to what is usually believed, the genuine cosmological constant has almost no scale dependence, even though
the zero point energies are taken into account. 
While this mechanism might not (fully) eliminate the finetuning problems related to the running gravitational parameters,
it changes the way they appear. In fact, even though the $k^4$ running is not visible in the effective Einstein equation,
there is still a $k^2$ running, via $M_k$, in the parameter $\Lambda_k^{\textrm{scb}}=\Lambda_k-\tfrac{3}{2}M_k$
which governs the curvature scale of selfconsistent backgrounds via \eqref{3.415}. (It is even conceivable that in a more
general truncation $F_k'(\chi_B, \chi_B)$ in \eqref{3.414} runs proportional to $k^4$.) Nevertheless, it will
certainly be worthwhile to reconsider the cosmological constant problem from this perspective \cite{inprep}.

From these remarks it should be clear that certain qualitatively important features of the gravitational RG flow can be 
uncovered only by going beyond the $\widehat{\Gamma}=0$ truncation. 
In the single metric truncations the fast
$k^4$-running is mis-attributed to the cosmological constant of the dynamical metric and gravitates therefore. 
Only a bimetric truncation can reveal that the RG running predominantly
affects  the action functional of the background metric only and that it does not occur in a source term for the dynamical one.

\subsection{The fixed points of the $\{g,\lambda,m\}$ subsystem}

In this subsection we find and analyze the fixed points $(g_*,\lambda_*,m_*)$ of the $\{g,\lambda,m\}$ subsystem of the 5-dimensional flow
\eqref{3.52}, that is, we search for common zeros of $\beta_g$, $\beta_{\lambda}$, and $\beta_m$. In the next subsection we shall then 
show how the fixed  points of the subsystem extend to fixed points of all 5 equations.

From \eqref{3.52a}, \eqref{3.52b}, \eqref{3.52c} with \eqref{3.53a} it is obvious that $\beta_g=\beta_{\lambda}=\beta_m=0$ admits a trivial
solution corresponding to a ``Gaussian" fixed point (GFP):
\begin{equation}\label{4.20}
g_*^{\rm{GFP}}=\lambda_*^{\rm{GFP}}=m_*^{\rm{GFP}}=0
\end{equation}

The GFP achieves $\beta_g\equiv (2+\eta_N)g=0$ by $g_*=0$. Alternatively one can satisfy this equation by setting $\eta_{N*}=-2$. Using
\eqref{3.53a} for the anomalous dimension this condition becomes, along with the other two equations $\beta_g=\beta_{\lambda}=0$,
\begin{subequations}\label{4.21}
\begin{align}
g_*\lambda_*^2 &= 3\pi\big( 1+m_* -2\lambda_*\big)^4
\label{4.21a}\\
g_*\lambda_* &=3\pi\big( 1+m_*\big)^3
\label{4.21b}\\
g_*\lambda_* &=-18\pi\, m_*\big( 1+m_*\big)^2
\label{4.21c}
\end{align}
\end{subequations}
These three conditions do indeed possess a solution, and this solution is unique. This non- Gaussian fixed point is located at
\begin{align}\label{4.22}
g_* &= \big(\frac{6}{7}\big)^2\, \frac{3\pi}{x_0}\approx 39.3
\nonumber\\
\lambda_* &=\frac{6}{7}x_0\approx 0.151
\nonumber\\
m_* &=-\frac{1}{7}\approx -0.143
\end{align}
Here $x_0\approx 0.176$ is the smaller one of the two real zeros\footnote{The analytic expression for $x_0 $
in terms of complicated radicals is not very illuminating. The second, obvious, zero of $\alpha(x)$ at $x=1$ does not lead to
a fixed point in the physical part of the theory space where $(1+m-2\lambda)>0$.}
presented by the function $\alpha(x)\equiv (1-2x)^4-x$.
%
%

The RG flow linearized about the NGFP is governed by the stability matrix $B_{ij}=\partial_j\beta_i(\mathbf{g}_*)$, $i,j=1, 2, 3$: 

\begin{equation}\label{4.23}
k\partial_k\; \mathbf{g}_i(k)= B_{ij}\Big(\mathbf{g}_j(k)-\mathbf{g}_{*j}\Big) 
\end{equation}
Here  we denote hte couplings collectivelly by $\mathbf{g}\equiv (\mathbf{g}_1, \mathbf{g}_2, \mathbf{g}_3)\equiv (g,\lambda, m)$.
Setting $t=\ln(k)$, the general solution of \eqref{4.23} reads
\begin{equation}\label{4.24}
\mathbf{g}_j(k)=\mathbf{g}_{*j}+ \sum_n r_n e^{i\alpha_n}\;e^{-\theta_n t}\;V_j^n
\end{equation}
Here $\{V^n,n=1,2,3 \}$ are the right eigenvectors of the stability matrix, with eigenvalues $-\theta_n$, and
$r_ne^{i\alpha_n}\equiv C_n$ are free constants of integration.
They can be complex except when $\theta_n$ happens to be real (then $\alpha_n=0$). Critical exponents with
$\textrm{Re}(\theta_n)>0$ correspond to relevant scaling fields. They grow when $k$ is lowered, i.e. they amount
to UV attractive directions.

For the NGFP  in \eqref{4.22} we have the following stability matrix:
\begin{equation}
{\mathbf{B}=
\left(
\begin{array}{ccc}
-2        &  -2175.02 &  566.518 \\
 0.0115 & -0.1747 & -1.0258 \\
  -0.0109 &  0.1654 & -3.6958
\end{array}
\right)}
\end{equation}
Diagonalizing $\mathbf{B}$ we find that the
linearized flow is governed by a pair of complex conjugate critical exponents
$\theta_1=\theta_2^*\equiv \theta'+i\theta''$ with $\theta'=1.071$ and $\theta''=5.535$, as well as
 a positive real one, $\theta_3=3.72805$. 
The nonzero imaginary part $\theta''\neq 0$ implies that the RG trajectories are spirals near the NGFP.
As $\theta'$ and $\theta_3$ are positive, the NGFP is UV-attractive in all three directions, i.e. all three scaling fields are relevant.
The  right eigenvectors associated with the eigenvalues $-\theta_1=-\theta_2^*$ and $-\theta_3$ are respectively, 
\begin{align}
V^1= &\big(1 \, , -0.00064 -  i\, 0.00213\, , -0.000821 + i\, 0.0016\big)=(V^2)^*\nonumber\\
V^3= & \big(0.989\,, 0.0376\,, 0.141\big)
\end{align}
We can rewrite eq. \eqref{4.23} as:
\begin{equation}
\mathbf{g}_j(k)=\mathbf{g}_{*j} +C_3 V_j^3 e^{-\theta_3 t}+ [V_j' \cos(\alpha-\theta''t)+V_j''\sin(\alpha-\theta''t)]\;e^{-\theta't}
\end{equation}
Here, $V'=\textrm{Re}[V^1]$ and  $V''=\textrm{Im}[V^1]$.
\begin{figure}[h!]\label{fig1}
\centering
\begin{tabular}{ll}
\bf{(a)} & \bf{(b)} \\
\includegraphics[width=7.3cm,height=5cm]{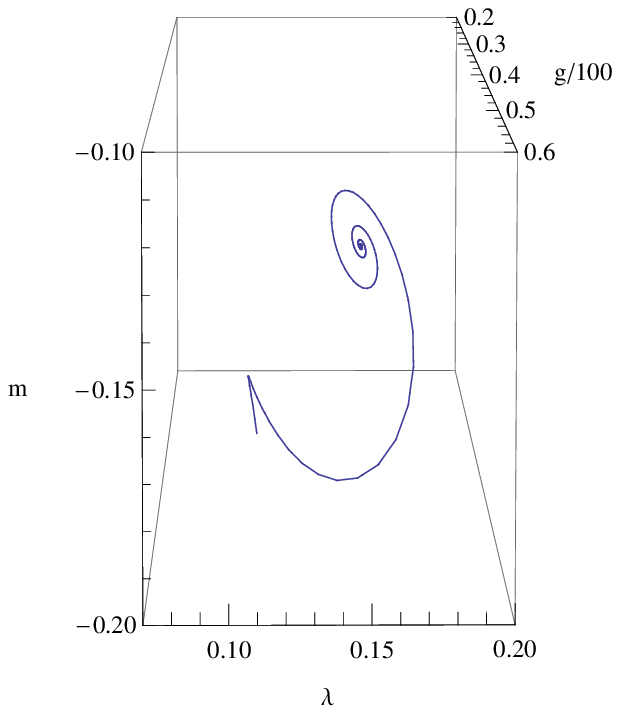} & \includegraphics[width=7.3cm,height=5cm]{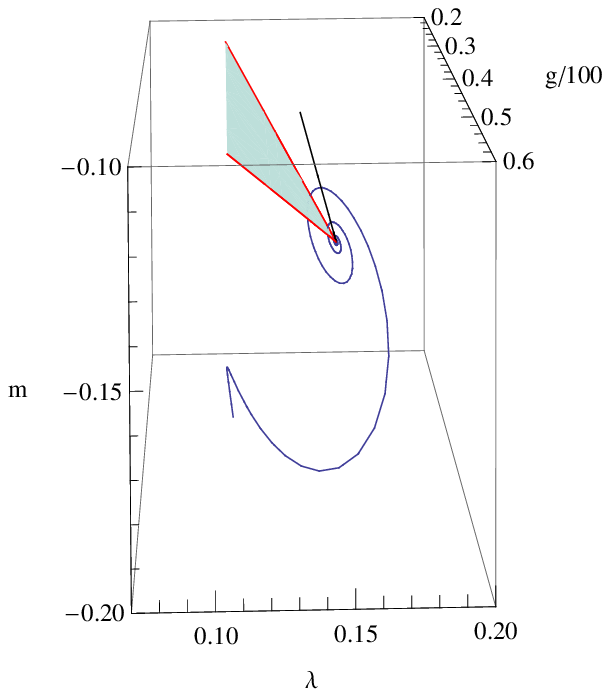} 
\end{tabular}
\caption{\small{The left figure shows a typical RG trajectory of the $\{g, \lambda, m\}$ subsystem which spirals into the NGFP for
$k\rightarrow \infty$. The right figure indicates in addition the eigendirections of the stability matrix. The shaded area is spanned by the vectors 
$V'$ and $V''$, and the straight line is in the direction of $V_3$. Since $\theta_3$ is large compared to $\theta'$, the spirals are slightly tilted relative to the $V'$-$V''-$ plane.}}
\end{figure}

 In Fig.1(a) we show a typical trajectory which spirals into the NGFP as $t\rightarrow \infty$. It was found by solving the full nonlinear 
$\{g, \lambda, m\}$ system numerically.
The figure 1(b) shows also the plane spanned by the vectors $V'$ and $V''$ at the NGFP. This plane  coincides almost with the plane
where the spirals form. Since $V^3$ is not coplanar to $V'$ and $V''$, it  shifts the trajectories away from the plane spanned by $V'$ and $V''$.

Figure 2 shows the three projections of various numerical solutions of the system \eqref{3.52}; their initial points 
are in the vicinity of the NGFP. 
 \begin{figure}\label{fig2}
\centering
\begin{tabular}{lll}
\bf{(a)} & \bf{(b)}& \bf{(c)}\\
\includegraphics[width=4.7cm,height=5cm]{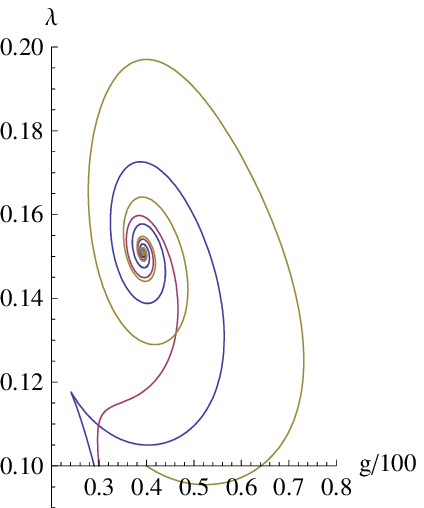}& \includegraphics[width=4.7cm,height=5cm]{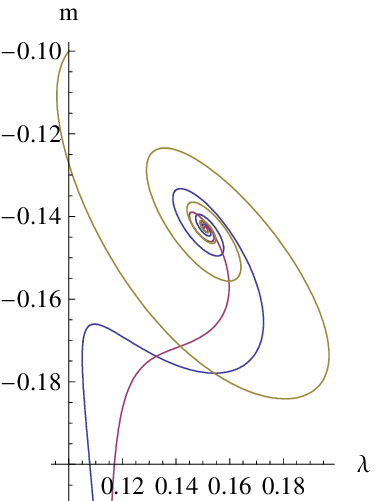} &
\includegraphics[width=4.7cm,height=5cm]{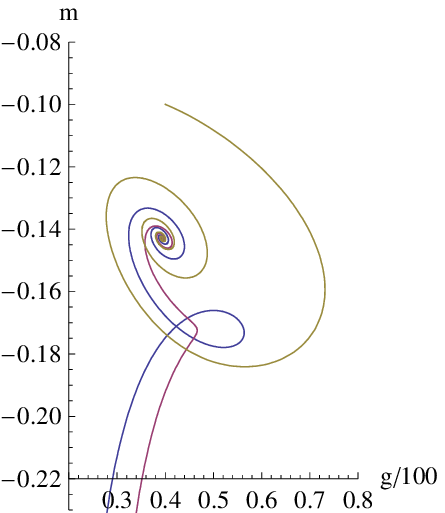}\\
\end{tabular}
\caption{\small{The plots show three different projections of the same RG trajectories, namely onto the $g$-$\lambda-$plane (a), the 
$\lambda$-$m-$plane (b), and  $g$-$m-$plane (c), respectively. }}
\end{figure}

Summarizing this analysis we can say that adding the $\phi^2\,\chi^2_B$-monomial to the CREH truncation has lead to a new relevant scaling field,
thus increasing the dimensionality of the UV critical surface of the NGFP by one unit. Since this dimensionality is related to the degree
of predictivity an asymptotically safe theory can have, we see that a proper understanding of the background dependence of $\Gamma_*$ is important also from this point of view.

\subsection{The fixed points of the 5 dimensional flow}

Next we discuss whether and how the GFP and NGFP of the subsystem generalize to fix points of all five flow equations \eqref{3.52}.

Let us begin with the GFP in the $\{g,\lambda,m\}$ subsystem. Inserting $g_*=\lambda_*=m_*=0$ into the remaining RG equations 
\eqref{3.52d} and \eqref{3.52e}  we obtain 
\begin{subequations}\label{4.40}
\begin{align}
k\partial_k\; g_k^B &=\beta_g^B=\big[  2+\eta_N^B(0,0,0;g_k^B)  \big]\,g_k^B
\label{4.40a}\\
k\partial_k\;\lambda_k^B &=\beta_{\lambda}^B=\big[\eta_N^B(0,0,0;g_k^B)-2\big]\,\lambda_k^B+\frac{g_k^B}{4\pi}
\label{4.40b}
\end{align}
\end{subequations}
with $\eta_N^B(0,0,0;g_k^B)=g_k^B/(12\pi)$. These equations admit two fixed points. The first one, denoted ``G-G-FP", achieves $\beta_g^B=0$
by $g^B_*=0$, the second one,  ``G-NG-FP", by $\eta_{N*}^B=-2$. Their respective locations are
\begin{align}
\textrm{G-G-FP:}\qquad  &g_*^B=0,\qquad\lambda_*^B=0
\label{4.41}\\
\textrm{G-NG-FP:}\qquad  &g_*^B=-(24\pi),\qquad\lambda_*^B=-\frac{3}{2}
\label{4.42}
\end{align}
Here and in the following the first label ``G" (or ``NG", later on) indicates whether  $\beta_g=0$ results from $g_*=0$ (``Gaussian")  or 
from $\eta_{N*}=-2$ (``non-Gaussian"). Likewise the second ``G" or ``NG" means that  $\beta_g^B=0$ is achieved via $g^B_*=0$ or 
 $\eta_{N*}^B=-2$, respectively.

Trying to extend the NGFP of the subsystem, we are led to consider
\begin{subequations}\label{4.43}
\begin{align}
k\partial_k\; g^B_k &=\beta_g^B=\Big[  2+\eta_N^B((g_*,\lambda_*,m_*)^{\rm{NGFP}};g_k^B)  \Big]\,g^B_k
\label{4.43a}\\
k\partial_k\;\lambda^B_k &=\beta_{\lambda}^B=\Big[\eta_N^B((g_*,\lambda_*,m_*)^{\rm{NGFP}};g_k^B)-2\Big]\,\lambda_k^B+
\frac{g_k^B}{3\pi}\frac{1}{1+m_*^{\rm{NGFP}}}
\label{4.43b}
\end{align}
\end{subequations}
Using the explicit formula for $\eta_N^B$ one finds that these  beta functions, too, possess two common zeros, one with 
$g^B_*=0$, the other with $\eta_{N*}^B=-2$:
\begin{align}
\textrm{NG-G-FP:}\qquad & g_*^B=0,\qquad\lambda_*^B=0
\label{4.44}\\
\textrm{NG-NG-FP:} \qquad& g_*^B=-\frac{108\pi}{49\xi_0}\approx -25.2,\qquad\lambda_*^B=-\frac{3}{14\xi_0}\approx -0.779
\label{4.45}
\end{align}
Here we abbreviated $\xi_0\equiv x_0+\tfrac{1}{84}(13-54x_0)(1-2x_0)^{-2}\approx 0.275$. Table 1 contains a summary of the four
fixed points we found.
%

\begin{table}
\begin{center}
\begin{tabular}{|c|c|ccccc|cc|}
\hline
Fixed Point & FP in the Subsystem & $g_*$ &  $\lambda_*$ & $m_*$ & $g_*^B$ & $\lambda_*^B$ & $g_*\lambda_*$ & $g_*^B\lambda_*^B$\\
\hline
  G-G-FP &  GFP & 0    & 0     & 0              & 0        & 0             & 0    & 0 \\
 G-NG-FP &  GFP & 0    & 0     & 0              & $-24\pi$ & $-\frac{3}{2}$& 0    & $36\pi$ \\
 NG-G-FP & NGFP & 39.3 & 0.151 & $-\frac{1}{7}$ & 0        & 0             & 5.93 & 0  \\
NG-NG-FP & NGFP & 39.3 & 0.151 & $-\frac{1}{7}$ & -25.2    & -0.779        & 5.93 & 19.6 \\
\hline
\end{tabular}
\end{center}
\caption{\small The table displays the coordinates of the fixed points in the 5-dimensional RG flow. The related fixed point in 
the subsystem, and the $g\lambda$ products are also given.}
\end{table}

Thus we may conclude that in the generalized truncation with $\widehat{\Gamma}\neq0$, too, the toy model of conformally reduced gravity 
possesses a non-trivial fixed point. Actually  there are two of them now, the NG-G-FP and the NG-NG-FP, but they differ only in their
$g^B_*$ and $\lambda_*^B$ values which do not affect the dynamical metric. Leaving infrared issues aside,
an asymptotic safety construction based upon either of these fixed points is conceivable. 
We consider this an encouraging result which suggests that  also full fledged QEG should continue to have a NGFP when the truncation is generalized
correspondingly.

While it is gratifying to see that the qualitative features of the flow, in particular near the fixed point, did not change, a
superficial glance at the results might convey the impression of considerable quantitative changes.
For instance, for the presumably universal quantity $g_*\lambda_*$ we obtained $g_*\lambda_*\approx 5.93$.
This figure has to be compared to $(g_*\lambda_*)_{\rm{CREH}}\approx 1.3$ found in \cite{creh1} with the simple CREH truncation. 
This change  is quite large in the following sense. In 
full QEG, the variations of $g_*\lambda_*$ under changes of $\mathcal{R}_k$ or of $\overline{\Gamma}_k$ within  the $\widehat{\Gamma}=0$ 
class of truncations  were at the $10\%$ level only. 
However, we must emphasize that it was clear from the outset that the toy model is much less stable under variations of $\mathcal{R}_k$ or
of the truncation than the full fledged QEG \cite{creh1,creh2, crehroberto}. And indeed, the changes of the critical exponents caused by the 
$\widehat{\Gamma}\neq 0$ modification are not larger than the variations within the $\widehat{\Gamma}=0$ scheme, comparing the 
``kin" and ``pot" versions of the CREH truncation \cite{creh1}, for example. It is plausible to expect that in full QEG the inclusion
of the leading bimetric invariants  has a smaller impact on the results. It is clear, though, that at a certain point the
$\widehat{\Gamma}\neq 0$ terms will be more 
 important than a further refinement of the truncations in the $\overline{\Gamma}_k$-sector.

The changes of the fixed point data are mainly due to the misinterpretation of background terms in the action
as dynamical ones which is unavoidable in a  $\widehat{\Gamma}=0$ truncation. The present truncation instead has no comparable 
``contamination" of the $\phi$-terms by $\chi_B$-contributions.

In this context it is interesting to note that the couplings in the ``purely background" sector have no reason to be  numerically
small compared to the others in the dynamical sector. As an extreme example, let us compare the two cosmological constants at the NG-NG-FP. 
There the prefactors of $k^4\int\sqrt{g}$ and $k^4\int \sqrt{\bar{g}}$, respectively, are $\lambda_*/g_*\approx0.38\times10^{-2}$ and 
$\lambda_*^B/g_*^B\approx 3.1\times10^{-2}$. Obviously the background term is about 10 times larger than the ``genuine" one.
The prefactors of the respective Einstein-Hilbert terms are of the same order of magnitude though.

\subsection{Testing the reliability of the truncation}

We mentioned already that the $\delta \Gamma_k/\delta\bar{g}$-equation \eqref{1.22} and likewise the $\delta \Gamma_k/\delta\chi_B$-equation
\eqref{3.23} in the toy model are exact relations, resulting from the same functional integral as the corresponding FRGE. 
Therefore exact solutions to the FRGE automatically satisfy those relations exactly, while approximate solutions of the FRGE, if they
are to be considered reliable, must satisfy them within the same degree of precision as the flow equation. In this section we use \eqref{3.23}, 
restricted to the potential, in order to assess the reliability of the five-parameter truncation used above. This relation is 
instantaneous with respect to the RG time, it does not couple different scales $k$, and so it can be checked at a single point of  theory space.
In particular we can test how well it is satisfied at the five fixed points found above.

This analysis is somewhat technical and the details are relegated to Appendix B. Here we only summarize the main results.

Analyzing \eqref{3.23} together with the fixed point condition on the still infinite dimensional theory space of arbitrary dimensionless
potentials $Y(\varphi,b)$ we find that the fixed point potential at the Gaussian fixed point is a pure $\varphi^2$ monomial, without any $b$
admixture. The non Gaussian fixed points instead have a strong, probably
predominant  $b$- component in their potential $Y_*(\varphi,b)$. Its precise form depends on the details of the $\mathcal{R}_k$. For equal
fields one has always $Y_*(\varphi,b=\varphi)\propto \varphi^4$.

Turning to the three-parameter potential ansatz which retains only the monomials $\phi^4$, $\phi^2\chi^2_B $ and $\chi_B^4$, with coefficients
 $\Lambda_k$, $M_k$, and $\Lambda_k^B$, respectively, we find that this truncation satisfies the $\delta \Gamma_k/\delta\chi_B$-equation
only if $\Lambda_k$ and $M_k$ are small compared to all  other relevant scales, and if $\Lambda_k^B/G_k^B\propto k^4$. This is a selection criterion 
for solutions of the projected FRGE. By picking appropriate initial conditions and possibly restricting the $k$ interval one certainly can find
consistent segments of RG trajectories. 

It is quite remarkable that the $\delta \Gamma_k/\delta\chi_B$-equation alone
(without the flow equation!) tells us that the genuine cosmological constant should be small in this truncation, and that its background 
counterpart has the notorious  $k^4$ dependence, well known from summing zero point energies.

Finally, checking how well the 4 fixed points of Table 1 satisfy the $\delta \Gamma_k/\delta\chi_B$-equation we find
that the G-G-FP and the G-NG-FP satisfy it exactly. The NG-G-FP and the NG-NG-FP are not exactly consistent, but the errors are still surprisingly
small (as compared to the large change of $g_*\lambda_*$ relative to its CREH value, say).

In summary we can say that the instantaneous $\chi_B$-equation for the potential does not hint at any severe inconsistency of the 
truncation which would matter at the qualitative level. For  the quantitative details we refer to Appendix B.


\section{Summary and conclusion}

Our  discussion started from the principle of ``background independence" which any satisfactory theory of quantum gravity should
respect. This requirement can be met in two complementary ways:
either one constructs the theory without using a background at all, or one does introduce some background, as a technical tool and for 
mathematical convenience, but shows that no prediction of the theory depends on it. Aiming at the quantization of gravity along the lines of
asymptotic safety we employed the gravitational average action approach here. It implements the second one of the above options by 
introducing an unspecified background metric $\bar{g}_{\mu\nu}$. In the computation of the average action this metric is kept completely
arbitrary, and in fact it gets promoted to a second argument of $\Gamma_k$, equally important as the dynamical metric $g_{\mu\nu}$.

In particular  the background metric is essential when it comes to devising a covariant and ``background independent" coarse graining scheme
for the gravitational field itself. The FRGE describing the scale dependence of the  average action necessarily operates on a theory
space consisting of functionals $\Gamma_k[g_{\mu\nu},\bar{g}_{\mu\nu},\cdots ]$ which depend on two metrics. Therefore this approach
to the quantization of gravity naturally has a certain bimetric character.

In the earlier applications of the FRGE approach \cite{mr}-\cite{je1} its bimetric nature had not yet been explored in concrete computations.
If one denotes the average action with the gauge fixing and ghost terms pulled out by $\overline{\Gamma}_k[g]+ \widehat{\Gamma}_k[g,\bar{g}]$
where, by definition, $\widehat{\Gamma}_k[g,g]=0$, then all truncations used so far set\footnote{Up to a wave function renormalization, at most.}
$\widehat{\Gamma}_k=0$. Projecting the RG flow on
$\overline{\Gamma}_k$ alone in deducing the beta functions from the exact FRGE certain $\bar{g}_{\mu\nu}$-contributions of 
the functional traces are interpreted as $g_{\mu\nu}$-contributions. Clearly one would like to estimate the error caused by this mis-attribution, 
which requires retaining $\widehat{\Gamma}_k$-type terms in the truncation ansatz. A second motivation for doing this is that
$\widehat{\Gamma}_k$ must be known in order to set up the effective field equations.

In full fledged QEG the computation of flows in  bimetric truncations is a formidable task.
For this reason we started in this paper the analysis of such truncations within the simpler setting of conformally reduced gravity,
a toy model which shares many properties with QEG, at least at the level of the  Einstein-Hilbert truncation.

After preparing the stage in Section 2, we introduced a special bimetric truncation for the conformally reduced model in Section 3. It involves two 
separate Einstein-Hilbert terms with constants, $G_k,\Lambda_k$ and $G_k^B,\Lambda_k^B$, respectively, for two conformally
flat metrics $g_{\mu\nu}=\phi^2\delta_{\mu\nu}$ and $\bar{g}_{\mu\nu}=\chi_B^2\delta_{\mu\nu}$, as well as a nonderivative mixed term
$\propto M_k$. From the point of view of the dynamical field $\phi$ the truncation ansatz is essentially a standard scalar 
$\phi^4$  action, with $\phi^4$ coupling $\Lambda_k$, and mass square $M_k$.
In Section 4 we  discussed the RG flow which follows from this truncation ansatz when a coarse graining operator
$\mathcal{R}_k$ is used which complies with the principle of ``background independence".
The main results can be summarized as follows.

\noindent
{\bf (A)} Identifying  $\bar{g}_{\mu\nu}$ with a fixed flat background metric, i.e. setting $\chi_B=$const, the EAA is found to be exactly
that of a standard scalar  field theory. We recover, for instance, the well known logarithmic running of $\Lambda_k$  and quadratic running
of $M_k$.

\noindent
{\bf (B)} The background cosmological constant $\Lambda_k^B$, i.e. the coefficient of $\sqrt{\bar{g}}=\chi_B^4$, has a very fast RG 
running proportional to $k^4$. We find that, in the EAA framework, {\it the  sum over zero point energies
contributes to the running of $\Lambda_k^B$, rather than to the scale dependence of the genuine cosmological constant $\Lambda_k$}, the coefficient of 
$\sqrt{g}=\phi^4$. As a result, according to the effective Einstein equations for finite $k$, it does not cause spacetime to curve.
This should be relevant to the cosmological constant problem.

\noindent
{\bf (C)} In the ``background independent" version of the earlier calculations of beta functions in the toy model  which do not disentangle $\bar{g}_{\mu\nu}$
and $g_{\mu\nu}$-contributions one sets $\bar{g}_{\mu\nu}=g_{\mu\nu}$ or $\chi_B=\phi$ at a certain stage of the calculation.
This turns the dependence of the regulator on the background metric into a dependence on the dynamical field :
$\mathcal{R}_k[\bar{g}_{\mu\nu}]\rightarrow\mathcal{R}_k[{g}_{\mu\nu}] $. The beta functions one obtains in this way are quite different 
from those of an ordinary scalar matter field on a non-dynamical flat Euclidean space; the latter beta functions are obtained using
$\mathcal{R}_k[{\delta}_{\mu\nu}]$.

 Since in the  calculation which disentangles  $\bar{g}_{\mu\nu}$
and $g_{\mu\nu}$-contributions the background metric is always kept unspecified, and neither put equal to $\delta_{\mu\nu}$ nor to
 $g_{\mu\nu}$, {\it the difference in the conceptual status of the ``background independent" and the ``rigid background" $\beta$ functions  
disappears completely}. 

In a certain sense, the new treatment combines both of the old calculations into one.  In the enlarged
theory space of the fully ``background independent" setting, with  $\widehat{\Gamma}_k\neq 0$, there are certain couplings whose
RG running reflects what happens in the rigid background treatment; the logarithmic running of the true cosmological constant $\Gamma_k$
is an example. But there are also couplings which are of no interest in the matter theory context, the prime example being $\Lambda_k^B$.

\noindent
{\bf (D)} In the generalized truncation we found nontrivial fixed points which suggest that the toy model is asymptotically safe.
Quantitatively, the $\widehat{\Gamma}_k\neq 0$  terms cause considerable changes of the fixed point  parameters, though.

\noindent
{\bf (E)} Thanks to the simplicity of the toy model it was possible to evaluate the instantaneous 
$\delta \Gamma_k/\delta\bar{g}_{\mu\nu}$-equation which governs the $\bar{g}_{\mu\nu}$-dependence of the EAA, and to use it for checking
the quality of the truncation used. Remarkably, this equation by itself, together with the specific truncation ansatz, is sufficient to predict that 
 the genuine cosmological constant is very small, and that $\Lambda_k^B/G_k^B$  runs proportional
to $k^4$.

\noindent
{\bf (F)} Finally we mention an issue we have not discussed yet. In \cite{elisa1} the information contained in the $k\rightarrow\infty$
limit of the EAA was used to construct a well defined regularized functional integral and a corresponding bare action for asymptotically
 safe theories. When applied to the conformally reduced Einstein-Hilbert truncation, without disentangling $\phi$ and $\chi_B$,
 this construction yields a bare 
action with a $\phi^4\ln{\phi}$ potential, while the effective potential  of the nontrivial fixed point is a 
simple $\phi^4$ term. Clearly it is somewhat unusual to find the Coleman-Weinberg potential $\phi^4\ln{\phi}$, nonanalytic in 
$\phi$, at the bare level. In Appendix C we show that also this somewhat strange phenomenon is an artifact of the truncations which do not
disentangle $\phi$ and $\chi_B$ contributions. In a bimetric truncation, the term which gets added in going from the effective to the bare fixed
point action is essentially $\chi_B^4\ln{\chi_B}$, and not  $\phi^4\ln{\phi}$.

\bigskip

Our discussion of the bimetric truncations in conformally reduced gravity should have made it clear that it is certainly worthwhile, and in fact
necessary to perform a similar analysis in full fledged QEG. 
Because of the better stability properties of the full theory it is plausible to expect that the impact of the $\widehat{\Gamma}\neq 0$
terms is smaller than in the toy model though. Those superior stability properties of the full gravity theory are, at least
partially, due to the fact that it contains two completely different types of interactions, both of which independently 
drive the formation of a NGFP: the self interactions of the
helicity-2 field coming from the $\sqrt{g}\, R$-term, and the vertices due to the cosmological constant term $\sqrt{g}\propto \phi^4$.
The toy model retains only the latter type.

Including $\widehat{\Gamma}\neq 0$ terms in  full QEG will require a considerable technical effort, however.
In particular new efficient strategies for the computation of functional traces  must be developed for this purpose.
It would also be interesting to reanalyze the nontrivial fixed points found in higher dimensional Yang-Mills theory theory
\cite{nonabavact, giesymfp} and the nonlinear sigma model \cite{nonlinsig} in the light of the above results.

\bigskip
{\bf Acknowledgements:} We would like to thank J. Pawlowski, R. Percacci, and O. Rosten for helpful discussions.

\vspace{2cm}

\appendix
\section*{Appendix}
\section{Derivation of the beta functions}
In this appendix we derive the flow equation for an arbitrary bimetric potential $F_k(\phi,\chi_B)$, the partial
differential equation \eqref{3.47}, as well as the beta functions of the five-parameter truncation displayed
in \eqref{3.52}. The starting point of all calculations is \eqref{3.46} which we shall project 
in various ways. We restrict ourselves to the ${R}^4$ topology and set $\widehat{g}_{\mu\nu}=\delta_{\mu\nu}$
throughout. All final results in particular those given in the main part of the paper are for the optimized shape function
\cite{opt}.

\subsection{The partial differential equation for $F_k(\phi,\chi_B)$ } 

In order to ``project out" the term $\propto \partial_k\;F_k$ from \eqref{3.46} we insert  $x$-independent configurations
$\phi,\chi_B=\rm{const}$ on both sides. In this case it is advantageous to evaluate the functional trace in the eigenbasis of
$\widehat{\square}\equiv \delta^{\mu\nu}\partial_{\mu}\partial_{\nu}$. Since, for $\chi_B=\rm{const}$, the Laplacians 
 $\overline{\square}$ and $\widehat{\square}$ are related by $\overline{\square}=\widehat{\square}/\chi^2_B$, and $\bar{R}=0$, the
resulting trace in eq.\eqref{3.46} is of the form ${\rm{Tr}}H(-\widehat{\square})$ which is simply 
 ${\rm{Tr}}H(-\widehat{\square})=(2\pi)^{-4}\int{\rm{d}}^4x\int{\rm{d}}^4p\;H(p^2)=2v_4\int{\rm{d}}^4x\int_0^{\infty}{\rm{d}}y \,y \;H(y)$ 
with $v_4=1/(32\pi^2)$. Upon changing the variable of integration from $y\equiv p^2$ to $z\equiv y/(k^2\chi^2_B)$ we find
\begin{align}\label{a.1}
\Big(-\frac{3}{4\pi G_k}\Big) & \Big[ k\partial_k\;-\eta_N\Big]  F_k(\phi,\chi_B)=  
\nonumber\\
{} & 2v_4\; (k\chi_B)^6
\int_0^{\infty}{\rm{d}}z\; z\;\frac{\big[R^{(0)}(z)-z{R^{(0)}}'(z) \big]-\frac{1}{2}\eta_NR^{(0)}(z)}{k^2\chi^2_B\;\big(z+R^{(0)}(z)\big)
+ F^{''}_{k}(\phi,\chi_B)}
\end{align}
This is as far as one can go for a generic function $R^{(0)}(z)$. In the following we shall employ the optimized shape function \cite{opt}
\begin{equation}\label{a.2}
R^{(0)}(z)=(1-z)\theta(1-z)
\end{equation}
which has the advantage that the $z$-integral in \eqref{a.1} can be performed in closed form. Using \eqref{a.2} in \eqref{a.1} a 
trivial calculation brings us to the final result given in eq.\eqref{3.47} of the main text.

\subsection{Projecting on the 3-parameter potential}

Next we investigate what happens when we restrict the functional form of $F_k(\phi,\chi_B)$ to the 3-parameter ansatz
\eqref{3.412} which contains the monomials $\phi^4$, $\phi^2\chi^2_B$ and $\chi_B^4$ only.
Again  we face the problem that the vector field which governs the RG flow on the big space of functions $F_k(\phi,\chi_B)$
 is not a priori tangent to the 3-dimensional subspace spanned by $\phi^4$, $\phi^2\chi^2_B$ and $\chi_B^4$. To get a flow on the
subspace, governed by a vector field tangent to it, a ``projection" has to be devised.

In order to streamline the notation we set
\begin{align}\label{a.3}
A_k\equiv -4\lambda_k, & \qquad B_k\equiv m_k, & C_k\equiv -\frac{1}{6}\lambda_k^B(g_k/g_k^B)
\end{align}
so that the 3-parameter potential \eqref{3.412} assumes the form
\begin{equation}\label{a.4}
F_k(\phi,\chi_B)=k^2 \Big[    \frac{A_k}{4!}\;\phi^4 +  \frac{B_k}{2} \;\phi^2\,\chi^2_B +  C_k  \; \chi_B^4        \Big]
\end{equation}
If we insert \eqref{a.4} straightforwardly on both sides of the RG equation for $F_k(\phi,\chi_B)$, eq.\eqref{3.47}, the result is
\begin{equation}\label{a.5}
\Big(k\partial_k\; +2-\eta_N\Big)\Big[    \frac{A_k}{4!}\phi^4 +  \frac{B_k}{2} \phi^2\chi^2_B +  C_k   \chi_B^4        \Big]=
-\frac{g_k}{24\pi}\Big(1-\frac{\eta_N}{6}\Big)\cdot Q(\phi,\chi_B)
\end{equation}
with the abbreviation
\begin{equation}\label{a.6}
 Q(\phi,\chi_B)\equiv \frac{\chi^6_B}{(1+B_k)\chi^2_B+\frac{1}{2}A_k\phi^2}
\end{equation}
In order to find the scale derivatives of the 3 couplings we must analyze the  various ``corners" of the $\phi$-$\chi_B$ plane
separately; they allow for different approximations (expansions) of the function $Q$.

If $\phi \gg \chi_B$ we may expand $Q(\phi,\chi_B)$ in powers of $(\chi_B/\phi)\ll 1$, yielding
\begin{equation}\label{a.7}
 Q(\phi,\chi_B)=\Big(\frac{2}{A_k}\Big)\frac{\chi_B^6}{\phi^2}\; 
\Big[ 1-2\frac{1+B_k}{A_k}\;\frac{\chi^2_B}{\phi^2} +4\Big(\frac{1+B_k}{A_k} \Big)^2 \;\frac{\chi_B^4}{\phi^4}+\cdots\Big]
\end{equation}
Obviously none of the of the terms $\propto {\chi_B^{\alpha}}/{\phi^{\beta}}$ appearing in this expansion matches those 
on the LHS of \eqref{a.5}. The conclusion is that, within this truncation, the RHS of \eqref{a.5} is equivalent to zero 
so that the couplings nave no nontrivial running. (Note that upon  returning 
to dimensionful parameters, and dividing them by the factor $G_k$ included in the
$\Gamma_k$ ansatz \eqref{3.40}, the ``+2" and the ``$-\eta_N$"  disappear from the first bracket of \eqref{a.5};
dimensionful parameters in $F_k/G_k$ have no $k$-dependence therefore.)

Conversely, when $\phi\ll \chi_B$ we can expand in $(\phi/\chi_B)\ll 1$, with the result
\begin{equation}\label{a.8}
 Q(\phi,\chi_B)= \frac{\chi_B^4}{1+B_k}-\frac{A_k}{2(1+B_k)^2}\;\chi^2_B\phi^2+ \frac{A_k^2}{4(1+B_k)^3}\;\phi^4 +\mathcal{O}
\big(\phi^6/\chi^2_B \big)
\end{equation}
We observe that the first 3 terms of the power series expansion in $\phi/\chi_B$ are precisely those retained in the truncation ansatz, and
the higher order terms are negligible when $ \phi\ll \chi_B$. As a consequence, equating the coefficients of $\phi^4$, $\phi^2\chi^2_B$ and
$\chi_B^4$ on both sides of \eqref{a.5} implies nontrivial RG equations:
\begin{align}\label{a.10} 
\frac{1}{4!}\Big( k\partial_k\; +2-\eta_N  \Big)A_k=& -\frac{g_k}{24\pi}\;\Big(1-\frac{\eta_N}{6}\Big)\;\frac{A_k^2}{4(1+B_k)^3}
\nonumber\\
\frac{1}{2}\Big( k\partial_k\; +2-\eta_N  \Big)B_k=& \frac{g_k}{24\pi}\;\Big(1-\frac{\eta_N}{6}\Big)\;\frac{A_k}{2(1+B_k)^2}
\nonumber\\
\Big( k\partial_k\; +2-\eta_N  \Big)C_k=& -\frac{g_k}{24\pi}\;\Big(1-\frac{\eta_N}{6}\Big)\;\frac{1}{1+B_k}
\end{align}
After returning to the $(\lambda,m,\lambda^B)$ variables the relations  \eqref{a.10} become exactly the equations
\eqref{3.52b}, \eqref{3.52c}, and \eqref{3.52e} presented in the main text.

To summarize: Depending on whether $(\phi/\chi_B)$ is large or small the RG equations on the 3-dimensional truncation subspace assume
different forms; they are given by, respectively, \eqref{a.10} and a set of similar equations with zero on their RHS. Loosely speaking,
different 3-dimensional RG flows correspond to different ``projections" of the flow on the larger theory space. Specifying
a truncation involves not only picking a set of monomials retained in the ansatz, it also involves specifying a projection.
Even after having fixed the functional form of $\Gamma_k[\phi,\chi_B]$, different regions of field space might still require
different projections, that is, different beta functions, for a reliable approximation.

In the present paper we employ only the beta functions corresponding to the case of small $\phi/\chi_B$. One of the reasons is that 
we would like to establish the connection between the gravity theory and a standard scalar theory. In the perturbative
quantization of the latter one sets $\chi_B\equiv1$ and makes a power series (or rather polynomial) ansatz for $F_k(\phi)$, leading to 
RG equations similar to \eqref{a.10}. Another reason is that we would like to have a simple model capable of describing 
the phase of unbroken diffeomorphism invariance in which the expectation value of the metric can vanish. This corresponds
to $g_{\mu\nu}=\phi^2\delta_{\mu\nu}\rightarrow 0$ at fixed, finite $\bar{g}_{\mu\nu}=\chi^2_B\delta_{\mu\nu}$, so that
$\phi/\chi_B\rightarrow 0$ in this situation.

\subsection{The anomalous dimension $\eta_N$}

Next we calculate the anomalous dimension $\eta_N$ for the 3 parameter form of $F_k$. To obtain an explicit formula for $\eta_N$ we
use eq.\eqref{3.46} for a constant background field, together with an $x$-dependent $\phi(x)=\chi_B+\bar{f}(x)$. As we
plan to perform the trace in the $\widehat{\square}$ eigenbasis we also insert $\overline{\square}=\widehat{\square}/\chi^2_B$ again:
\begin{align}\label{a.11}
\frac{3}{8\pi G_k}\;\eta_N & \int\textrm{d}^4x\, \bar{f}(x)(-\widehat{\square}) \bar{f}(x)=\nonumber\\
 {} & \chi^2_B k^2 \;\textrm{Tr}\Bigg[ \frac{\Big(1-\frac{\eta_N}{2}\Big)R^{(0)}\Big( \frac{-\widehat{\square}}{\chi^2_Bk^2}\Big)
-\Big( \frac{-\widehat{\square}}{\chi^2_B k^2}\Big){R^{(0)}}'\Big( \frac{-\widehat{\square}}{\chi^2_B k^2}\Big)}{-\widehat{\square}+ 
\chi^2_B k^2 %
R^{(0)}\Big(\frac{-\widehat{\square}}{\chi^2_B k^2}\Big) + M_k\chi^2_B-2\Lambda_k\Big(\chi_B+\bar{f}(x)\Big)^2
}\Bigg]
\end{align}
Consistent with the truncation we must perform a derivative expansion of the trace, thereby retaining only the term which is of 
second order both in $\bar{f}$ and the number of derivatives. As it turns out, this term is automatically independent of
$\chi_B$, as is the LHS of \eqref{a.11}. Hence we have the same monomial on both sides and obtain  a nonzero $\eta_N$

Actually we do not have to redo the explicit derivative expansion here since almost the same trace as in \eqref{a.11} has been
evaluated in earlier investigations of the toy model in ref.\cite{creh1}. (See in particular the computation of $\eta^{\rm{kin}}$
and the Appendix A there.)
The only modification is the $M_k\chi^2_B$ term in the denominator which is easy to take care of.
From the results of \cite{creh1} we can easily read off the answer for the anomalous dimension:
\begin{equation}\label{a.12}
\eta_N(g,\lambda,m)=-\frac{8}{3\pi}g\lambda^2\;\frac{\widehat{\Sigma}_4(m-2\lambda)}{1+\frac{4}{3\pi}g\lambda^2\;\widetilde{\Sigma}_4(m-2\lambda)}
\end{equation}
Here $\widehat{\Sigma}_4(w)$ and $\widetilde{\Sigma}_4(w)$ are threshold functions defined in \cite{creh1}. They can be evaluated for any
$R^{(0)}$. For the optimized shape function they read
\begin{align}\label{a.13}
\widehat{\Sigma}_4(w)=\frac{1}{4}\frac{1}{(1+w)^4}, \quad &\quad \widetilde{\Sigma}_4(w)=0
\end{align}
When we insert \eqref{a.13} into \eqref{a.12} we obtain the final result \eqref{3.53a} given in the main text.

\subsection{The anomalous dimension $\eta_N^B$} 

In order to compute the anomalous dimension related to $G_k^B$ we could in principle evaluate \eqref{3.46} for an 
$x$-dependent $\chi_B$ and constant $\phi$; then only the $\chi_B\widehat{\square}\chi_B$ term would contribute
on the LHS. Actually it turns out that the trace to be evaluated simplifies
if, instead, we insert two equal, nonconstant fields: $\chi_B(x)=\phi(x)$. Then also the $\phi\widehat{\square}\phi$ term
contributes, but its prefactor $\eta_N$ is known already. For the 3 parameter potential this leads to:
\begin{align}\label{a.14}
\frac{1}{16\pi} & \Bigg(\frac{\eta_N}{G_k}+ \frac{\eta_N^B}{G_k^B}\Bigg) \int\textrm{d}^4x\,\sqrt{\bar{g}}R(\bar{g}) =
  k^2 {\rm{Tr}}\Bigg[ 
\Bigg\{\Big(1-\frac{\eta_N}{2}\Big)R^{(0)}\Big( \frac{-\overline{\square}}{k^2}\Big) 
\nonumber\\
{} & -\Big( \frac{-\overline{\square}}{ k^2}\Big){R^{(0)}}'\Big( \frac{-\overline{\square}}{ k^2}\Big) \Bigg\}
\Bigg(-\overline{\square}+  k^2
R^{(0)}\Big(\frac{-\overline{\square}}{ k^2}\Big) +\frac{1}{6}\bar{R}(x)+ M_k-2\Lambda_k
\Bigg)^{-1}
\Bigg]
\end{align}
On the LHS we used that $\int\,\sqrt{\bar{g}}R(\bar{g})=6\int\textrm{d}^4x\delta^{\mu\nu}\partial_{\mu}\chi_B\partial_{\nu}\chi_B$
for the conformally flat metric $\bar{g}_{\mu\nu}=\chi^2_B\delta_{\mu\nu}$. To find  $\eta_N^B$ we now
perform a covariant derivative expansion of the trace in \eqref{a.14}, retaining only the Einstein-Hilbert term. This trace
depends on $\bar{g}_{\mu\nu}$ in a twofold way: via the Laplace-Beltrami operator pertaining to  $\bar{g}_{\mu\nu}$, 
$\overline{\square}$, and via the curvature scalar $\bar{R}(x)$. It is most convenient to take care of the first dependence
by heat kernel methods \cite{mr} and to expand out the second explicitly. Actually a very similar calculation has already been performed
in \cite{creh1}, albeit in a different context and with a different interpretation. (See in particular the derivation of
$\eta_N^{\rm{pot}}$ in \cite{creh1}.) Performing obvious replacements the trace term of interest can be read off from the
results in \cite{creh1}:
\begin{equation}\label{a.15}
\frac{\eta_N}{g}+\frac{\eta_N^B}{g^B}= B_1(\lambda-m/2)+\eta_NB_2(\lambda-m/2)
\end{equation}
The functions  $B_1$ and $B_2$ are the same as in \cite{creh1}, but the $M_k\phi^2\chi^2_B$ term shifts their argument.
For the optimized shape function we have explicitly
\begin{align}\label{a.16}
B_1(\lambda-m/2)= & \frac{1}{12\pi}\;\frac{1+2m-4\lambda}{(1+m-2\lambda)^2}
\nonumber\\
B_2(\lambda-m/2)= & -\frac{1}{36\pi}\;\frac{1+ \frac{3}{2}m-3\lambda}{(1+m-2\lambda)^2}
\end{align}
When we solve eq.\eqref{a.15} for $\eta_N^B$ we obtain precisely the formula \eqref{3.53b} given in the main text.


\section{Testing the reliability of the truncation}

In this appendix we employ the $\delta\Gamma_k/\delta\chi_B$ equation \eqref{3.23} in order to assess the reliability of the truncation
used in this paper. At the exact level, a RG trajectory $k\mapsto \Gamma_k$ which solves the FRGE is automatically a solution of the 
 $\delta\Gamma_k/\delta\chi_B$ equation. When we perform approximations this is no longer the case, and the degree with which the latter equation
is satisfied can be used to judge the quality of an approximate solution to the FRGE. In particular this can be done when the FRGE and the 
 $\delta\Gamma_k/\delta\chi_B$  equation are projected onto some subspace.

For the general $F_k$ ansatz \eqref{3.40}, a calculation similar to the derivation of $k\partial_k\;F_k$ shows that the
 $\delta\Gamma_k/\delta\chi_B$  equation implies for the dimensionful and dimensionles potential, respectively,
\begin{align}
\chi_B\;\frac{\partial}{\partial\chi_B}F_k(\phi,\chi_B) & = -\frac{G_k}{24\pi}\;
\frac{\chi_B^6k^6}{\chi^2_Bk^2+\partial_{\phi}^2\;F_k(\phi,\chi_B)}
\label{c.1}\\
b\;\frac{\partial}{\partial b}Y_k(\varphi,b) & = -\frac{g_k}{24\pi}\;
\frac{b^6}{b^2+\partial_{\varphi}^2\;Y_k(\varphi,b)}
\label{c.2}
\end{align}
These relations hold true for the optimized shape function. Obviously they have a similar structure
as the truncated flow equation. They are instantaneous, however; they contain no derivative with respect to $k$. In
particular, for a trajectory with UV fixed point, eq.\eqref{c.2} is valid at $k=\infty$, i.e. it constrains the fixed point
potential $Y_*(\varphi,b)$.

\subsection{Fixed point potentials at the general $Y_k$ level}

 To the extent some approximate RG trajectory $k\mapsto Y_k(\varphi,b)$ is (approximately) consistent with the
projected FRGE and the $\delta\Gamma_k/\delta\chi_B$  equation, any linear combination of these two equations will also be approximately
satisfied by this trajectory. For instance, we may combine \eqref{3.51} and \eqref{c.2} in such a way that the complicated term involving 
$\partial_{\varphi}^2Y_k(\varphi,b)$ disappears:
\begin{equation}\label{c.3}
\Big[k\partial_k\; -(2+\eta_N)+\varphi\partial_{\varphi}+\frac{1}{6}\eta_N\;b\partial_b\Big]Y_k(\varphi,b)=0
\end{equation}
The term $\propto\eta_N/6$ in \eqref{c.3} stems from the factor $(1-\eta_N/6)$ on the RHS of the flow  equation \eqref{3.51}. It is due
to the factor $1/G_k$ in the normalization of the $\mathcal{R}_k$, see eq.\eqref{3.22}. This prefactor contributes a term $\propto\eta_N$
to $\partial_k\;\mathcal{R}_k$.

It is instructive to consider \eqref{c.3} at a fixed point, $\partial_k\;Y_k=0$. At the GFP we have $g_*=0$ and $\eta_N=0$. Hence \eqref{c.2}
tells us that $b\partial_bY_*(\varphi,b)=0$ and \eqref{c.3} yields $[\varphi\partial_{\varphi}-2]Y_*(\varphi,b)=0$. Taken together
these relations imply that
\begin{equation}\label{c.4}
Y_*^{\rm{GFP}}(\varphi,b)=c\;\varphi^2
\end{equation}
where $c$ is a constant (which can be fixed as in \cite{creh2}).
Note that the fixed point potential \eqref{c.4} depends only on the {\it dynamical} field $\varphi$, not on the background.

At a NGFP, instead, one has $\eta_{N*}=-2$, and \eqref{c.3} boils down to
\begin{equation}\label{c.5}
\Big[\varphi\partial_{\varphi}-\frac{1}{3}b\partial_b\Big]Y_*^{\rm{NGFP}}(\varphi,b)=0
\end{equation}
This equation would imply that $Y_*^{\rm{NGFP}} $ is a function of a single variable, $b^3\varphi$.
However, there exists no exact fixed point solution to the flow equation \eqref{3.51} of the form $Y_*^{\rm{NGFP}}(\varphi, b)=f(b^3\varphi)$. 
This reflects the fact that we are dealing with approximate equations, and for consistency must be content with potentials which solve
these equations only approximately.

Nevertheless the qualitative conclusion we can draw from \eqref{c.5} is that, at a NGFP, the fixed point potential  $Y_*^{\rm{NGFP}} $ 
is likely to involve both the dynamical field $\varphi$ and the background $b$. Moreover, there is a clear trend for the background
field to be predominant. In fact if we change the normalization of $\mathcal{R}_k$ and replace $1/G_k$ in its prefactor by a constant,
the $\eta_N/6$ term disappears from the flow equation and from \eqref{c.3}. As a consequence, eq.\eqref{c.5} gets replaced by
$\varphi\partial_{\varphi}Y_*(\varphi,b)=0$. This equation alone would then imply that $Y_*^{\rm{NGFP}} $ is a function of $b$ alone!
Clearly, as we said above, we are dealing with  approximate solutions to approximate equations so that this result cannot be taken 
at face value. But still it highlights the importance of the background field for the structure of the fixed point action.

\subsection{The 3-parameter potential ansatz}

Now  we perform a further projection of the $\delta\Gamma_k/\delta\chi_B$ equation \eqref{c.1} namely on potentials of the form
\eqref{3.412}. It is then found to imply
\begin{equation}\label{c.6}
\Lambda_k=0,\qquad M_k=0,\qquad\Lambda_k^B=\frac{1}{16\pi}G_k^Bk^4
\end{equation}
Again, these conditions should not be taken at face value, but a reliable solution of the projected FRGE should satisfy them 
approximately. Hence the basic message of \eqref{c.6} is clear: {\it The $3$-parameter potential ansatz which retains only the monomials
$\phi^4, \phi^2\chi^2_B$, and $\chi_B^4$ can be consistent only when the genuine cosmological constant $\Lambda_k$ is small
 compared to all other relevant scales and its background counterpart scales as $G_k^Bk^4$. }

It is quite remarkable that the $\delta\Gamma_k/\delta\chi_B$ equation alone is sufficient to predict that $\Lambda_k^B\propto k^4$
when $\eta_N^B=0$, the result which obtains by summing zero point energies. 

It is perhaps even more interesting and possibly
relevant to the cosmological constant problem that {\it the very structure of the potential ansatz, and therefore of the
(conformally reduced) Einstein-Hilbert action in particular, favors a vanishing or at most small cosmological constant $\Lambda_k$.}

The conditions \eqref{c.6} read in dimensionless form
\begin{equation}\label{c.7}
\lambda_k=0,\qquad m_k=0,\qquad\lambda_k^B=\frac{1}{16\pi}g_k^B
\end{equation}
We interpret the first two of them as the requirement that $\lambda_k$ and $m_k$ should be much smaller than unity, or at least
significantly smaller than other typical couplings.

\subsection{The fixed points from the 3-parameter potential}

Finally we check to what extent the four   fixed points of the $\{g, \lambda, m; g^B, \lambda^B\}$ system in Table 1 are in accord
with \eqref{c.7}. One finds that the G-G-FP and the G-NG-FP satisfy \eqref{c.7} exactly. 

The NG-G-FP has the coordinates
$$(g, \lambda, m; g^B, \lambda^B)_*= (39.3, 0.151, -0.143; 0, 0)$$ Obviously $\lambda_*$ and $m_*$ are ``anomalously small"
at this fixed point (compared to $g_*$, say) and in particular much smaller that unity. This is indeed what \eqref{c.7} requires,
and its third relation is satisfied exactly even.

The coordinates of the NG-NG-FP are $$(g, \lambda, m; g^B, \lambda^B)_*= (39.3, 0.151, -0.143; -25.2, -0.779)$$
The $\lambda_*$ and $m_*$ values are the same as above, and from $g^B_*$ and $\lambda^B_*$ we get $g^B_*/ \lambda^B_*=(16\pi)/1.56$.
According to the last relation of \eqref{c.7} this ratio should be equal to $16\pi$. Our result is off by a factor of $1.56$.
We interpret this number which is still remarkably close to unity as an indication that our treatment of the 
NG-NG-FP is at least qualitatively correct, and that this fixed point is presumably not an artifact of the truncation.


\section{From the effective to the \\
bare fixed point action}

The limit $k\rightarrow\infty$ of the EAA is closely related to the {\it bare} action which appears in the integrand of the
underlying functional integral. In ref.\cite{elisa1} we explained how, after having introduced an UV regulator at scale\footnote{ There
should be no confusion between the UV cutoff $\mathbf{\Lambda}$ and the cosmological constant.}
$\mathbf{\Lambda}$  into the integral, the corresponding bare action $S_{\mathbf{\Lambda}}$ relates to $\Gamma_k$ if one requires
that the integral with $S_{\mathbf{\Lambda}}$ reproduces the prescribed EAA in the limit $\mathbf{\Lambda}\rightarrow\infty$.
Under certain conditions a fixed point $\Gamma_*$ at the effective level implies a fixed point $S_*$ 
for the bare action, and one can derive an explicit functional differential equation for $S_*$
in dependence on
$\Gamma_*$. For the specific UV regulator used in \cite{elisa1} (``finite mode" regularization) it reads
\begin{equation}\label{5.1}
\Gamma_{\mathbf{\Lambda}}[\phi,\chi_B]- S_{\mathbf{\Lambda}}[\phi,\chi_B] =
 \frac{1}{2}{\rm{Tr}}\Bigg\{\theta(\overline{\square}+\mathbf{\Lambda}^2)\ln \Big[S_{\mathbf{\Lambda}}^{(2)}[\phi,\chi_B]+\mathcal{R}_{\mathbf{\Lambda}}[\chi_B]      \Big]  \Bigg\}
\end{equation}
Here $\Gamma_{\mathbf{\Lambda}}$ stands for $\Gamma_k$ evaluated at $k=\mathbf{\Lambda}$, and $S_{\mathbf{\Lambda}}^{(2)}$ denotes the Hessian
of $S_{\mathbf{\Lambda}}$ with respect to $\phi$. Given $\Gamma_k$ for  $k\rightarrow\infty$, eq.\eqref{5.1} may be used to find 
$S_{\mathbf{\Lambda}}$ in the limit  $\mathbf{\Lambda}\rightarrow\infty$. (See \cite{elisa1} for further details.)

In \cite{elisa1} we applied this algorithm to conformally reduced gravity, making a local potential ansatz, without extra $\chi_B$ dependence
$(\widehat{\Gamma}=0)$, for both $\Gamma_k$ and $S_{\mathbf{\Lambda}}$. The former  action contains the familiar dimensionless potential 
$Y_k(\varphi)$, the one in the latter is denoted $\check{Y}_{\mathbf{\Lambda}}(\varphi)$. For the ${R}^4$ topology the 
effective NGFP potential turned out to be a pure $\varphi^4$ monomial, $Y_*(\phi)\propto\varphi^4 $, while the corresponding bare 
fixed point potential was much more complicate, behaving for $\varphi\rightarrow \infty$ asymptotically as
\begin{equation}\label{5.11}
\check{Y}_*(\varphi)\approx \frac{\check{g}_*}{48\pi} \varphi^4\ln\varphi^2
\end{equation}
where $\check{g}_*$ is the bare Newton constant in $S_*$. While this potential is of the familiar Coleman-Weinberg from, it is
remarkable that here it appears as part of the bare action corresponding to a simple $\varphi^4$ monomial in the effective one.
Thus, compared to a standard scalar matter field theory, the situation is exactly inverted. Since a bare action nonanalytic in the
field is a somewhat unusual situation it is worthwhile to reconsider this issue with a bimetric truncation.

As we explained  in Appendix B, we cannot exclude the possibility that in a bimetric truncation the potential ${Y}_*(\varphi,b)$
in $\Gamma_*$ is a complicated, possibly nonanalytic function of both $\varphi$ and $b$. However, we can exclude the possibility
that the mere transition from the effective to the bare level generates  Coleman-Weinberg like nonanalytic terms. 

To see this in the simplest  setting possible we make an ansatz for $\Gamma_k$ in which the potential $F_k(\phi,\chi_B)$ does
have an extra $\chi_B$ dependence but not the kinetic term:
\begin{equation}\label{5.2}
\Gamma_k[\phi,\chi_B]=-\frac{3}{4\pi G_k}\int\,{\rm{d}}^4x\,\sqrt{\widehat{g}}\, \Big(\frac{1}{2}\widehat{g}^{\mu\nu}
\partial_{\mu}\phi\partial_{\nu}\phi +F_k(\phi,\chi_B)\Big)
\end{equation}
For the bare action we make an analogous ansatz, with a bare Newton constant $\check{G}_{\mathbf{\Lambda}}$ and a potential
$\check{F}_{\mathbf{\Lambda}}(\phi,\chi_B)$:
\begin{equation}\label{5.3}
S_{\mathbf{\Lambda}}[\phi,\chi_B]=-\frac{3}{4\pi \check{G}_{\mathbf{\Lambda}}}\int\,{\rm{d}}^4x\,\sqrt{\widehat{g}}\,
 \Big(\frac{1}{2}\widehat{g}^{\mu\nu}\partial_{\mu}\phi\partial_{\nu}\phi +\check{F}_{\mathbf{\Lambda}}(\phi,\chi_B)\Big)
\end{equation}
Now it is straightforward to insert \eqref{5.2}, at $k=\mathbf{\Lambda}$, and \eqref{5.3} into \eqref{5.1} and to project
out the nonderivative terms. In terms of the dimensionless quantities \eqref{3.48}, \eqref{3.49}, and
analogously defined bare ones the result reads
\begin{equation}\label{5.4}
\frac{Y_{\mathbf{\Lambda}}(\varphi, b)}{g_{\mathbf{\Lambda}}}-
\frac{\check{Y}_{\mathbf{\Lambda}}(\varphi, b)}{\check{g}_{\mathbf{\Lambda}}}=
-\frac{1}{48\pi}\;b^4\;\ln  \Bigg\{ \frac{b^2+\partial_{\varphi}^2\;\check{Y}_{\mathbf{\Lambda}}(\varphi, b)}{\check{g}_{\mathbf{\Lambda}}}  \Bigg\}
\end{equation}
This is a complicated differential equation for $\check{Y}_{\mathbf{\Lambda}}$ when $Y_{\mathbf{\Lambda}}(\varphi, b)$ is a given
solution of the FRGE, evaluated at $k=\mathbf{\Lambda}$. For $\mathbf{\Lambda}\rightarrow\infty$ it relates the fixed point
potentials $Y_*$ and $\check{Y}_*$.

Eq.\eqref{5.4} is strikingly different from its analog in the earlier $\widehat{\Gamma}=0$ calculation where the $\varphi$ and $b$ 
contributions were not disentangled. There we found:
\begin{equation}\label{5.5}
\frac{Y_{\mathbf{\Lambda}}(\varphi)}{g_{\mathbf{\Lambda}}}-
\frac{\check{Y}_{\mathbf{\Lambda}}(\varphi)}{\check{g}_{\mathbf{\Lambda}}}=
-\frac{1}{48\pi}\;\varphi^4\;\ln  \Bigg\{ \frac{\varphi^2+\partial_{\varphi}^2\;\check{Y}_{\mathbf{\Lambda}}(\varphi)}{\check{g}_{\mathbf{\Lambda}}}\Bigg\}
\end{equation}
Obviously the explicit $b$'s on the RHS of \eqref{5.4} are misinterpreted as $\varphi$'s in the simpler truncation leading to \eqref{5.5}.
The Coleman-Weinberg potential \eqref{5.1} is a direct consequence of this misinterpretation. At the NGFP the effective potential is
$Y_{\mathbf{\Lambda}}(\varphi)\propto \varphi^4$ for which one can easily show that \eqref{5.5} has the asymptotic solution \eqref{5.1}.

For the ''correct" equation \eqref{5.4} the situation is different: If $Y_{\mathbf{\Lambda}}(\varphi, b)$
is analytic in $\varphi$, so is the bare potential $\check{Y}_{\mathbf{\Lambda}}(\varphi, b)$. In this case the RHS of \eqref{5.4}
has the structure $b^4\ln [b^2+ a_0(b)+a_1(b)\varphi+a_2(b)\varphi^2+\cdots]$. Since, as always, $b\neq 0$, this expression admits
 a power series expansion in $\varphi$. So we see that when the {\it effective} fixed point potential is a power series in $\varphi$,
the {\it bare} potential is a power series, too.

\end{document}